\documentclass[11pt,twoside]{article}

\usepackage{geometry}                
\usepackage{graphicx}
\usepackage{amsmath,amssymb, amsthm,epsfig,bm}
\usepackage{setspace}
\usepackage{epstopdf}
\usepackage{psfrag}
\usepackage{color,soul}
\usepackage{verbatim}
\usepackage{algorithm} 
\usepackage{multirow}
\usepackage{natbib}
\usepackage{bibentry}
\usepackage{xr}
\usepackage{mathrsfs}
\usepackage{cases}
\usepackage[colorlinks=true,citecolor=blue,linkcolor=blue]{hyperref}
\usepackage[shortlabels]{enumitem} 

\usepackage[utf8]{inputenc} 
\usepackage[T1]{fontenc}    
\usepackage{hyperref}       
\usepackage{url}            
\usepackage{booktabs}       
\usepackage{amsfonts}       
\usepackage{nicefrac}       
\usepackage{microtype}      
\usepackage{xcolor}         

\usepackage{amsmath}
 \usepackage[noend]{algpseudocode}
\usepackage{bbm}
\usepackage{subcaption}
\usepackage{comment}
\usepackage{enumitem}
\usepackage{microtype}
\usepackage{array, booktabs}

\usepackage{caption}
\usepackage{mwe}

\newcommand{\titleshort}{Distributed causal graph learning}
\newcommand{\authorshort}{Ye, Amini, and Zhou}

\setlist{noitemsep, topsep=0pt}
\bibpunct{(}{)}{;}{a}{}{,} 
\usepackage[english]{babel}
\usepackage[utf8]{inputenc}
\usepackage{fancyhdr}
 
\usepackage{bm}

\newcommand\bu{\overline b}
\newcommand\bl{\underline b}
\newcommand\ball{\mathbb B}
\newcommand\beh{\widehat \beta}

\newcommand\grad{\nabla}
\newcommand\gt{\widetilde g}
\newcommand\nt{\widetilde n}

\DeclareMathOperator*\argmin{argmin}

\newcommand\pih{\widehat \pi}
\newcommand\betah{\widehat \beta}
\newcommand\lt{\widetilde \ell}

\newcommand\PA{\text{PA}}

\newcommand\prox{\text{prox}}

\newcommand\Ic{\mathcal I}

\newcommand\lamin{\lambda_{\min}}

\newcommand{\f}{\frac}

\newcommand{\R}{\mathbb{R}}

\newcommand{\eps}{\varepsilon}

\newcommand{\norm}[1]{\|#1\|}

\newcommand{\summ}[2]{\sum_{#1 = 1}^{#2}}

\newcommand{\calC}{\mathcal{C}}
\newcommand{\calD}{\mathcal{D}}

\newcommand{\calG}{\mathcal{G}}

\newcommand{\calI}{\mathcal{I}}

\newcommand{\calM}{\mathcal{M}}
\newcommand{\calN}{\mathcal{N}}
\newcommand{\calO}{\mathcal{O}}
\newcommand{\calP}{\mathcal{P}}

\newcolumntype{H}{>{\setbox0=\hbox\bgroup}c<{\egroup}@{}} 

\newtheorem{theorem}{Theorem}
\newtheorem{definition}{Definition}

\newtheorem{prop}{Proposition}
\newtheorem{remark}{Remark}

\usepackage{xspace} 
\newcommand{\MATLAB}{\textsc{Matlab}\xspace}

\newcommand{\ip}[1]{\langle #1\rangle}
\newcommand{\reals}{\mathbb{R}}

\newcommand\mnorm[1]{|\!|\!|#1|\!|\!|}
\newcommand\fnorm[1]{\mnorm{#1}_F}

\newcommand{\pr}{\mathbb{P}}
\newcommand{\ex}{\mathbb{E}}

\usepackage{array,multirow}

\makeatletter
\newcommand*{\addFileDependency}[1]{
  \typeout{(#1)}
  \@addtofilelist{#1}
  \IfFileExists{#1}{}{\typeout{No file #1.}}
}
\makeatother
\newcommand*{\myexternaldocument}[1]{%
    \externaldocument{#1}%
    \addFileDependency{#1.tex}%
    \addFileDependency{#1.aux}%
}
\myexternaldocument{darls_supp}

\onehalfspace

\begin{document}

\title{Distributed Learning of Generalized Linear Causal Networks}
\author{Qiaoling Ye, Arash A. Amini and Qing Zhou\thanks{Department of Statistics, University of California, Los Angeles. Email: yeqiaoling@g.ucla.edu, aaamini@stat.ucla.edu, zhou@stat.ucla.edu}}
\date{}
\maketitle

\begin{abstract}
We consider the task of learning causal structures from data stored on multiple machines, and propose a novel structure learning method called distributed annealing on regularized likelihood score (DARLS) to solve this problem. We model causal structures by a directed acyclic graph that is parameterized with generalized linear models, so that our method is applicable to various types of data.
To obtain a high-scoring causal graph, DARLS simulates an annealing process to search over the space of topological sorts, where the optimal graphical structure compatible with a sort is found by a distributed optimization method. This distributed optimization relies on multiple rounds of communication between local and central machines to estimate the optimal structure. We establish its convergence to a global optimizer of the overall score that is computed on all data across local machines. To the best of our knowledge, DARLS is the first distributed method for learning causal graphs with such theoretical guarantees.
Through extensive simulation studies, DARLS has shown competing performance against existing methods on distributed data, and achieved comparable structure learning accuracy and test-data likelihood with competing methods applied to pooled data across all local machines.
In a real-world application for modeling protein-DNA binding networks with distributed ChIP-Sequencing data, DARLS also exhibits higher predictive power than other methods, demonstrating a great advantage in estimating causal networks from distributed data. 

\end{abstract}


\section{Introduction}

Causal diagrams are mathematical models for cause-effect relationships among variables~\citep{Pearl95}, with applications in education~\citep{DW99, Hill12}, economics~\citep{Imbens04}, and epidemiology~\citep{M06}, etc.
A causal diagram is represented by a {\em directed acyclic graph} (DAG), where edges encode causal effects among variables (nodes). 
Graphical models associated with DAGs are known as {\em Bayesian networks} (BNs). 
The probability density of a set of random variables $\{X_1, \ldots, X_p\}$ in a BN factorizes as
\begin{equation}\label{eq:bn_def}
	p(x_1, \ldots, x_p) = \prod_{j=1}^p p(x_j \mid \PA_{j} = pa_{j}), 
\end{equation}
where $\PA_{j} \subset \{X_1, \ldots, X_p\} \setminus \{X_j\}$ is the parent set of ${X_j}$ with $pa_{j}$ being its value.
Since randomized experiments are not always feasible, various approaches have been put forward to learn causal DAGs from observational data~\citep{Pearl09, Guo_etc_20}. 

In this work, we focus on learning causal DAGs from distributed data.
Distributed data storage has been used for privacy protection when managing the copious amount of data generated everyday by government agencies, research institutions, medical centers, technology companies, etc~\citep{Mehmood_etc_16}. Many of these organizations collect similar data, or data from the same population, and they often collaborate in social, scientific and business domains. For example, in 2021, Google and Apple were initiating an exposure notification system that tracks potential exposure to COVID-19 and a privacy-preserving analytics platform that collects health metrics~\citep{ENPA}. Also, small clinics often combine data to obtain statistically significant results since the individual datasets are otherwise too small. Such collaborations always require privacy disclosures stating that each institution cannot share the privately-hold sensitive data with the other parties. 
Even when data privacy is not a concern, merging data from multiple sources remains a difficult task. Since each local agency stores and manages data with different platforms, it is often time-consuming to extract, standardize and migrate the data. From a purely computational perspective, distributed computing often uses parallelism across local machines, resulting in substantial reduction in computation time and allowing us to scale inference procedures to massive datasets.

The widespread use of distributed data has lead to a research area whose goal is to adapt common statistical and machine learning tasks to the distributed setting. A straightforward idea for deriving a distributed version of any inference procedure is to form a global estimate by averaging the local estimates, an approach known as \emph{one-shot parameter averaging}. This approach, however, fails to obtain solutions with any desired level of suboptimality compared to the global estimate constructed based on all the data~\citep{ZWL10, SSZ14}. To overcome drawbacks of one-shot averaging, communication-efficient algorithms have been proposed that utilize multiple rounds of communication between local and central machines to generate a sequence of (global) estimates~\citep{ZDW13, SSZ14, JLY18, FGW19}.
Communication-efficient algorithms are particularly useful in  distributed optimization of multi-agent systems, such as electronic power systems, sensor networks and smart manufacturing \citep{Metc17, YANG19}.

Despite these methodological advances, learning causal DAGs from data distributed across independent machines is still a challenging task. A major difficulty is how to integrate local information to form a global causal graph that satisfies the acyclicity constraint. One approach is to iterate over local datasets (once) and then combine the local graphs or the local p-values to form a global graph~\citep{NY10, GJZ10, TWNA19}. However, there is no theoretical guarantee that aggregating local estimates, using this single-iteration approach, would lead to an estimate close to the global minimizer of the loss on the combined data. 
In this paper, we propose a score-based learning method that carries out multiple rounds of communication to estimate DAGs from distributed data. Our objective function is equivalent to a regularized log-likelihood of the overall data. To optimize it, the central machine proposes a candidate ordering $\pi$, where the score of $\pi$ is evaluated via communication with local machines over DAGs compatible with $\pi$. Then, the candidate sort $\pi$ is selected by simulated annealing. Because every DAG has at least one ordering, searching over the space of orderings ensures that the acyclicity constraint is always satisfied. We show that the convergence rate of our distributed estimate to the global one is $O(\log(n)/\sqrt{m})$ for a fixed true DAG, where $n$ is the total sample size across all local machines and $m$ is the smallest local sample size~ (Theorem~\ref{thm:conv:concrete}, Section~\ref{sec:convergence}). To the best of our knowledge, our approach is the first to learn causal graphs from distributed data with a theoretical guarantee of convergence to the global estimate as the sample size of the local machines grow.

Another contribution of our work is the use of {\em generalized linear models} (GLMs) for local conditional distributions in BNs, which brings several advantages to causal structure learning.  Our proposed GLM DAG model is a flexible family for various data types beyond linear Gaussian models (with equal variance) and multi-logit models~\citep{GFZ18, AGZ19}, and hence it can be applied to multiple domains. Additionally, most models in the GLM family lead to convex loss functions which facilitate the optimization task. The objective function in our distributed learning algorithm is equivalent to a regularized likelihood of the overall data, which has been shown to be effective in learning both continuous and discrete DAGs \citep{FZ13, AZ15, GFZ18, YAZ19}. Furthermore, we show that GLM DAG models lead to the identifiability of the underlying causal DAGs (Proposition~\ref{prop:multi-node-ident2}, Section~\ref{sec:ident}), while other common models, such as multinomial for discrete networks and Gaussian linear DAGs, are not identifiable in general~\citep{Pearl95, HGC95}. Under such identifiability, we establish the $\ell_2$-consistency of a global maximizer DAG of our regularized likelihood score (Theorem~\ref{prop:perm-const}, Section~\ref{sec:consistency}). 

The paper is organized as follows. Section~\ref{sec:gkn} defines the generalized linear DAG model and establishes some identifiability results under this model. In Section~\ref{sec:d-dag-learning}, we set up the optimization problem for learning causal graphs and develop the DARLS algorithm that combines simulated annealing to search over permutation space and a distributed optimization algorithm to optimize the network structure given an ordering. We then establish theoretical results for the convergence of the distributed optimization algorithm and estimation consistency of DARLS in Section~\ref{sec:theoretical}. Section~\ref{sec:numerical} consists of exhaustive simulation experiments, where we compare our method to existing ones using distributed data, test the robustness of DARLS against violations of its underlying model assumptions, and examine the accuracy loss and computational efficiency of distributed learning. We also apply the distributed learning methods to ChIP-sequencing data for modeling protein-DNA binding networks in Section~\ref{sec:real-data}. The paper is concluded with a discussion in Section~\ref{sec:conclude}. All proofs are relegated to the supplementary material.


\section{Generalized linear DAG models}
\label{sec:gkn} \label{sec:ident}

Denote by $x^j \in \R^{d_j}$ a realization of variable $X_j$, where $d_j = 1$ for a numerical $X_j$ and $d_j = r_j - 1$ for a categorical variable $X_j$ with $r_j$ classes, using the one-hot encoding. Let $\beta_{ij} \in \R^{d_i \times d_j}$ encode the influence of $X_i$ on $X_j$ and $\beta_{ij} = 0$ if $X_i \notin \PA_{j}$.
Put
\begin{align}\label{eq:x:betaj:def}
    \beta_j := [\beta_{0j}, \beta_{1j}, \ldots, \beta_{pj}] \in \R^{(d+1) \times d_j}, \quad x := [1, x^1, \ldots, x^p] \in \R^{d+1},
\end{align}
where $\beta_{0j} \in \R^{1 \times d_j}$ and $d = \summ i p d_i$. Here and elsewhere, $[x,y]$ denotes the vertical concatenation of two vectors or matrices $x$ and $y$. 
We define a {\em generalized linear DAG} (GLDAG) as the Bayesian network~\eqref{eq:bn_def} with conditional densities given by GLMs with canonical links, that is, 
\begin{align}\label{eq:expo-fam-def}
	p (x^j \mid pa_{j}, \beta_j ) &= \exp\left( \langle   \beta_j^\top x , x^j \rangle -  b_j (   \beta_j^\top x ) \right) + c_j (x^j), \quad j \in [p]
\end{align}
where $b_j$ and $c_j$ are both functions from $\reals^{d_j}$ to $\reals$. Note that $\beta_j^\top x=\sum_{i\in\PA_j} \beta_{ij}^\top x^i$ only depends on $pa_j$. GLDAG models allow for many common distributions via the choice of the log partition-function $b_j(\cdot)$. Examples include the Bernoulli distribution for $b_j(\theta) = \log(1 + e^\theta)$, constant-variance Gaussian for $b_j(\theta) = \theta^2/2$, Poisson {for} $b_j(\theta) = \exp(\theta)$, Gamma {for} $b_j(\theta) = -\log(-\theta)$ and the multinomial for $b_j(\theta) = \log\bigl(1 + \summ l {d_j} e^{\theta_l}\bigr)$.  Note that in the multinomial case $b_j(\cdot)$ is a multivariate function, operating on a vector $\theta = (\theta_l)$, in contrast to the other example for which $b_j(\cdot)$ is a scalar function. The Bernoulli and multinomial choices above give rise to logistic and multi-logit regression models for each node.

We collect all the parameters of model~\eqref{eq:expo-fam-def} in a matrix $\beta \in \R^{(d+1)\times d}$ which is obtained by horizontal concatenation of $\beta_j, j=1,\dots,p$, each as defined in~\eqref{eq:x:betaj:def}. We say that a GLDAG~\eqref{eq:expo-fam-def} is continuous if all the variables are continuous. Recall that in this case, $d_j = 1$ for all $j \in [p]$ and thus $\beta$ is a $(p+1) \times p$ matrix. We rewrite the log pdf of \eqref{eq:expo-fam-def}, in the continuous case, as
\begin{align}\label{eq:glm-sem-log-pdf}
    L(x; \beta) = \sum_{j=1}^p \big[ \log c_j(x^j) + x^j (\beta^\top x)_j - b_j\big((\beta^\top x)_j\big) \big],
\end{align}
where $\beta^\top x\in\R^{p}$ and $\beta_{ij} \neq 0$ if and only if $X_i \to X_j$. Next, we define identifiability of DAG models following~\cite{PMJS14-JMLR, HJMPS08, PB14}, and show that continuous GLDAGs are identifiable.

\begin{definition}[Identifiability]
Suppose we are given a joint distribution $L(X)=L(X_1,\ldots,X_p)$ that has been generated from an unknown GLDAG model~\eqref{eq:expo-fam-def} with a graph $\calG_0$. 
If the distribution $L(X)$ cannot be generated by any GLDAG model with a different graph $\calG \neq \calG_0$, then we say $\calG_0$ is identifiable from $L(X)$. 
\end{definition}

It is well-known that linear Gaussian DAGs and multinomial DAGs in general are not identifiable~\citep{Pearl95, HGC95}. In contrast, continuous GLDAG models~\eqref{eq:glm-sem-log-pdf} are identifiable under mild assumptions:

\begin{prop}\label{prop:multi-node-ident2}
Suppose the joint distribution $L(X)$ is defined by the log-pdf $L(x; \beta)$ with a DAG $\calG_0$ according to \eqref{eq:glm-sem-log-pdf} such that $\beta_{ij} \neq 0$ if and only if $i \in \PA_j$ in $\calG_0$.
If $L(x; \beta)$ is second-order differentiable with respect to $x$ and the first-order derivative of $b_j(\cdot)$ exists and is not constant for all $j$, then $\calG_0$ is identifiable from $L(X)$. 
\end{prop}

Proposition~\ref{prop:multi-node-ident2} establishes the identifiability of continuous GLDAG models \eqref{eq:glm-sem-log-pdf}, partially justifying our goal as to learn causal graphs. This result also expands the class of identifiable DAG models in the literature. DAGs generated from linear Gaussian structural equation models with equal variance can be fully identified~\citep{PB14}, which is a special case of the GLDAG models with $b_j(\theta) = \theta^2/2,~\forall j$. A different class of identifiable DAG models is the additive noise model, $X_j = f_j(\PA_j) + \eps_j$, assuming nonlinear $f_j$ and/or non-Gaussian $\eps_j$~\citep{lingam06, HJMPS08, PMJS14-JMLR}.

\section{Distributed DAG learning}
\label{sec:d-dag-learning}

In this section, we construct the objective function using distributed data and propose a simulated annealing search combined with an iterative optimization method to learn causal DAG structures. We start with the definition of topological sorts for DAGs. Given a {permutation} $\pi$ on $[p]:=\{1, \ldots, p\}$, we permute a vector $v = (v_1, \ldots, v_p)$ according to $\pi$ to obtain a relabeled vector $v_\pi = \left(v_{\pi(1)}, \ldots, v_{\pi(p)}\right)$. A {\em topological sort} of a DAG is a permutation of nodes such that if $a \in \PA_b$, then $a$ precedes $b$ in the order defined by $\pi$, denoted by $a \prec_\pi b$. By definition \eqref{eq:bn_def}, every DAG has at least one topological sort.

Let $\{x_h\}_{h=1}^n$ be an i.i.d. sample of size $n$ from model~\eqref{eq:expo-fam-def}. We also let $x_h^j$ represent the observed value of the $j$-th variable ($X_j$) in the $h$-th data point. Consider a subset $\Ic \subset [n]$. The normalized negative log-likelihood of the subsample $\{x_h\}_{h \in \Ic}$ is given, up to an additive constant, by
\begin{align}\label{eq:glm-loss-beta}
	\ell_\Ic(\beta) := 
	\frac1{|\Ic|} \sum_{h \in \Ic} \summ j p \bigl[b_j( \beta_j^\top x_h ) -   \ip{\beta_j^\top x_h, x_h^j}\bigr].
\end{align}
Note that in this notation, $\ell_{[n]}$ denotes the normalized negative log-likelihood of the entire sample of size $n$.

\subsection{Global objective function and annealing}
\label{sec:opt}

We consider the case that the overall data is stored on $K$ different servers, where each local machine $\calM_k$ holds its private data $\{x_h\}_{h \in \Ic_k}$ and communicates with a central machine $\calC$. Let $n_k = |\Ic_k|$ be the sample size in $\calM_k$ so that $\sum_{k=1}^K n_k = n$. The normalized negative log-likelihood based on the entire data can be decomposed as $\ell_{[n]}(\beta) = \sum_{k=1}^K \frac{n_k}{n}\ell_{\Ic_k}(\beta)$. Let $\calP$ be the set of all permutations on $[p]$ and $\calD(\pi) \subset \reals^{(d+1) \times d}$ the set of DAGs whose topological sorts are compatible with a permutation $\pi\in\calP$. Note that $\calD(\pi)$ is a linear subspace of $\reals^{(d+1) \times d}$. We ideally would like to estimate $\beta$ by minimizing a regularized loss function of the form
\begin{align} \label{eq:learn-bn-obj}
	\min_{\pi \in \calP} f(\pi), \quad \text{where} \quad f(\pi):= \min_{\beta \in \calD(\pi)} \summ k K  
	\frac{n_k}n \ell_{\calI_k}(\beta) + \rho(\beta),
\end{align}
and $\rho(\cdot)$ is an appropriate regularizer to promote sparsity in $\beta$. We call $f(\pi)$ the global objective function since it is defined using all data across local machines.

Recall that $\beta_{ij} \neq 0$ if and only if $i \in \PA_j$. To learn sparse DAGs, we apply group regularization of the form
\begin{equation}\label{eq:group-lasso}
    \rho(\beta) =  \lambda \sum_{i,j} \rho_g(\beta_{ij}),
\end{equation}
where $\rho_g(\cdot)$ is a nonnegative and nondecreasing group regularizer and $\lambda > 0$ is a tuning parameter. 
Restricted to $\mathcal D(\pi)$, the regularizer can be further simplfied to $\rho(\beta) = \lambda  \sum_j \sum_{i\prec_\pi j} \rho_g \left(\beta_{ij}\right)$. In this paper, we consider the Group Lasso (i.e., group $\ell_2$) penalty with the choice
\begin{equation} \label{eq:group-lasso-scalar}
    \rho_g (\beta_{ij}) = \fnorm{\beta_{ij}},
\end{equation}
where $\fnorm{\beta_{ij}}$ is the Frobenius norm of matrix $\beta_{ij}$. As a convex penalty and a natural extension of Lasso regularization, Group Lasso has demonstrated remarkable performance in grouped variable selection~\citep{YL07}.

To search over $\left(\pi \in \calP, \beta \in \calD(\pi)\right)$ with distributed data as in~\eqref{eq:learn-bn-obj}, we propose the {\em distributed annealing on regularized likelihood score} (DARLS) algorithm, which applies annealing strategies to search over the permutation space, coupled with a distributed optimization method. Such manner of joint optimization over the topological sort space and the DAG space has demonstrated great effectiveness in learning BNs; see for example, \cite{Larranaga1996, FK03, SCC15, YAZ19} and the references thereof. 

The main steps of DARLS are outlined in Algorithm~\ref{algo:general}. At each annealing iteration, a permutation $\pi^*$ is proposed based on current $\pih$ (line~\ref{alg:pi-prop}) and is accepted with probability according to simulated annealing given a decreasing temperature schedule. To compute the score of the optimal DAG structure for a given permutation, we use the {\em distributed optimization} approach outlined in Algorithm~\ref{algo:dist-iter-opt}, the details of which are discussed in~Section~\ref{sec:distr-opt} below.  This approach allows multiple rounds of communications between local and the central machines to update and synthesize information. Note that DARLS can be applied to any objective function as long as the gradient w.r.t. $\beta$ has a closed-form expression. 

\begin{algorithm}[t]
\caption{Distributed annealing on regularized likelihood score (DARLS).} \label{algo:general}
\smallskip
\hspace*{\algorithmicindent} \textbf{Input:} $\{x_h\}_{h=1}^n$ distributed over $K$ machines, $\pi_0$, a temperature schedule $\{T^{(i)}\}_{i = 0}^N$, $\tau$. \\
\hspace*{\algorithmicindent} \textbf{Output:} $\pih, \betah$. 
\begin{algorithmic}[1]
	\State Select tuning parameter $\lambda$ by BIC selection. \label{alg:bic} 
	\State $\pih \gets \pi_0$, compute $(\betah, f(\pih))$ by Algorithm~\ref{algo:dist-iter-opt}. \label{alg:dist-data}
	\For{$i = 0, \ldots, N$} 
		\State $T \gets T^{(i)}$.
		\State Central machine $\calC$ proposes $\pi^+$ by flipping a random interval (length up to $\tau$) in $\pih$. \label{alg:pi-prop}
		 \State Compute $(\beta^+, f(\pi^+))$ using Algorithm~\ref{algo:dist-iter-opt}. \label{alg:agl2}
		\State $\calC$ sets $(\pih, \betah, f(\pih)) \gets (\pi^+, \beta^+, f(\pi^+))$ with prob. $\min \left\{1, \exp\big({-}\frac1T[f(\pi^+)-f(\pih)]\big)\right\}$.\label{alg:next-step}
	\EndFor
	\State \textbf{end for} \label{algo1:end}
	\State Refine the causal structure implied by $\betah$. 
	\label{alg:refine-beta}
\end{algorithmic}
\end{algorithm}

\begin{algorithm}[t]
\caption{Distributed optimization to compute the global permutation score.}\label{algo:dist-iter-opt}
\hspace*{\algorithmicindent} \textbf{Input:} $\pi$, $\beta^{(0)}_\pi$, number of iteration $T$. \\
\hspace*{\algorithmicindent} \textbf{Output:}  $\betah_\pi, f(\pi)$. 
\begin{algorithmic}[1]
	 \State Central processor $\calC$ broadcasts $\pi$ to local machines $\{\calM_k\}_{k = 1}^K$.
	\For{$t = 0, 1, \ldots,  T-1$}
		\State  Each machine $\calM_k$ computes $\nabla \ell_{\calI_k}(\beta^{(t)}_\pi)$ and sends the result to $\calC$. 
		\State $\calC$ computes $\nabla \ell_{[n]}\big(\beta^{(t)}_\pi\big)= \f 1 n \sum_k    {n_k}  \nabla \ell_{\calI_k}\big(\beta^{(t)}_\pi\big)$ and broadcasts it to local machines.
		\label{algo:local-gradients}
		\State Each $\calM_k$ calculates the minimizer
		$\beta^{(t+1)}_{k,\pi} = \varphi_{k,\pi}\big(\beta^{(t)}\big)$ given by~\eqref{eq:local:update} and sends it to $\calC$. \label{algo:local-update}
		\label{algo:local-optimization}
		\State $\calC$ computes $\beta^{(t+1)}_\pi =  \f 1 n  \sum_k  {n_k} \beta^{(t+1)}_{k,\pi}$ and broadcasts it to local machines.
		\label{algo:io-beta-update}
		\EndFor
		\State \textbf{end for}
	\State Each $\calM_k$ reports $F_k^{(T)} : =n_k F_k\big(\beta^{(T)}_\pi\big)$ to $\calC$, and $\calC$ sets $\betah_\pi \gets \beta^{(T)}_\pi$ and  $f(\pi) \gets \f 1 n \sum_k  F_k^{(T)}$.
\end{algorithmic}
\end{algorithm}

\subsection{Local objective functions and distributed optimization}
\label{sec:distr-opt}

For any fixed $\pi$, we use distributed computing to evaluate $f(\pi)$. That is, instead of directly working with the objective function in~\eqref{eq:learn-bn-obj}, we rely on local versions of it to guide a distributed algorithm that divides the task of computing $f(\pi)$ among the $K$ local machines. In particular, we consider the local objective functions
\begin{align*}
	f_k(\pi) := \min_{\beta \in \calD(\pi)} F_k(\beta), 
	\quad \text{where} \quad 
	F_k(\beta) := \ell_{\calI_k}(\beta) + \rho(\beta).
\end{align*}
The global version~\eqref{eq:learn-bn-obj} can be rewritten as $f(\pi) = \min_{\beta \in \calD(\pi)} F(\beta)$ where $F(\beta) := \ell_{[n]}(\beta) + \rho(\beta)$. Typically, each of $F$ and $F_k$ is nonsmooth due to the presence of the regularizer $\rho$, but the difference $h_k := F_k - F = \ell_{\calI_k} - \ell_{[n]}$ is often smooth. The gradient of $h_k$ is used to guide iterations in each local machine. That is, given the current (global) estimate $\beta$, local machine $\calM_k$ performs the update
\begin{align}\label{eq:local:update}
    \varphi_{k,\pi}(\beta) := \argmin_{\xi \in \calD(\pi)} \big[\widetilde F_k(\xi) := F_k(\xi) - \ip{\nabla h_k(\beta), \xi} = F_k(\xi) - \ip{\grad \ell_{\calI_k}(\beta) - \grad \ell_{[n]}(\beta), \xi} \big].
\end{align}
The local regularized loss $F_k$ guided by $\grad h_k$, denoted by $\widetilde F_k$, is a first-order approximation to the global regularized loss $F$, up to an additive constant. Let $ \beta_\pi^{(t)}$ be the global estimate of the algorithm at iteration $t$. At the next iteration, $t+1$, we obtain local estimates $ \beta^{(t+1)}_{k,\pi} = \varphi_{k,\pi}(\beta^{(t)}_\pi)$ for $k=1,\dots,K$. These local estimates are then passed to the central machine $\calC$ to compute the next global estimate by averaging, i.e., $ \beta_\pi^{(t+1)} = \sum_{k=1}^K \frac{n_k}{n} \beta_{k,\pi}^{(t+1)}$. The main steps of this distributed optimization method are outlined in Algorithm~\ref{alg:dist-data}. 

The above approach is essentially a version of the DANE algorithm~\citep{ZDW13, SSZ14, JLY18, FGW19}. Note that to calculate local updates $\beta_{k,\pi}^{(t+1)}$ (line~\ref{algo:local-optimization}, Algorithm~\ref{alg:dist-data}), only the current global estimate $\beta_\pi^{(t)}$ and the global gradient $\nabla \ell_{[n]}(\beta_\pi^{(t)})$ need to be communicated to each local machine. In Section~\ref{sec:convergence}, we show that for a sufficiently large minimum sample size per machine, i.e. $\min_k n_k$, the sequence $\{\beta_\pi^{(t)}\}_{t \ge 0}$ thus produced will converge to a global minimizer $\widehat\beta_\pi$ of $F(\cdot)$ over $\calD(\pi)$.

Another piece in the distributed optimization is to compute the local update~\eqref{eq:local:update} (line~\ref{algo:local-update}, Algorithm~\ref{algo:dist-iter-opt}), for which we use the proximal gradient algorithm (Algorithm~\ref{algo:pg}). Given a current global estimate $\beta^{(t)}$, optimizing local objective $\widetilde F_k(\xi)$~\eqref{eq:local:update} is equivalent to
\begin{equation*}
    \min_{\xi \in \calD(\pi)} \ell_{\calI_k}(\xi) - \langle \nabla h_k (\beta^{(t)}), \xi \rangle + \rho(\xi).
\end{equation*}
Define $\lt_{\calI_k}(\xi) := \ell_{\calI_k}(\xi) - \langle \nabla h_k (\beta^{(t)}), \xi \rangle$, a surrogate for the global likelihood $\ell_{[n]}(\xi)$. 
To solve~\eqref{eq:local:update}, we use iterative proximal gradient descent. At each iteration, we minimize a quadratic approximation to $\lt_{\calI_k}$ around the current solution $\xi$, plus a regularization term, 
\begin{align}\label{eq:local:update:prox}
	\xi^+ := \argmin_{\xi' \in\calD(\pi)} \left[\lt_{\calI_k}(\xi) + 
	\ip{\grad \lt_{\calI_k}(\xi), \xi' - \xi } + \f 1 {2s} \fnorm{\xi' - \xi}^2 + \rho(\xi')\right],
\end{align}
where $s > 0$ plays the role of a step size and $\xi^+$ is our next estimate of the solution. Equivalently, the update \eqref{eq:local:update:prox} can be re-written as
\begin{align}\label{eq:proximal-grad}
\xi^+  = \text{prox}_{s\rho}\left( \xi - s \grad \lt_{\calI_k}(\xi)\right),
\end{align}
where $\text{prox}_{s\rho}(x) := \argmin_u \left(s \rho(u) + \f 1 2 \fnorm{x-u}^2\right)$ is a proximal operator applied to the scaled function $s\rho(\cdot)$. Equation~\eqref{eq:proximal-grad} is known as a {\em proximal gradient update}, and for our choice of the regularizer given by~\eqref{eq:group-lasso} and~\eqref{eq:group-lasso-scalar},  has the following closed-form expression:
\begin{align*}
	\left(\text{prox}_{s \rho} (x)\right)_{ij} = \left(1 - \f s { \fnorm{x_{ij}}}\right)_+ x_{ij}.
\end{align*}
This is often referred to as the \emph{block soft-thresholding} operator.
To determine the value of the step size $s$, we use backward line search, shrinking an initial value $s_0$ until a proper step size is found. We refer readers to \cite{NP13} for more details on the proximal algorithms.

\begin{algorithm}[t]
	\caption{Use the proximal gradient algorithm to compute local  permutation scores.}
	\label{algo:pg}
	\hspace*{\algorithmicindent} \textbf{Input:} $\{x_h\}_{h \in \calI_k}$, $\pi$, $\beta^{(t-1)}\in \calD(\pi)$, $\nabla h_k(\beta^{(t-1)})$, $s_0 > 0$, $\kappa \in (0,1)$, \texttt{max-iter}, \texttt{tol}. \\
	\hspace*{\algorithmicindent} \textbf{Output:}  $\beta_{k,\pi}^{(t)} $. 
	\begin{algorithmic}[1]
		\State {iter} $\gets 0$, err $\gets \infty$, $\xi \gets \beta^{(t-1)}.$
		\While{{iter} $< $ \texttt{max-iter} and {err} $>$ \texttt{tol}}
			\State $\nabla \lt_{\calI_k}(\xi) \gets \nabla \ell_{\calI_k}(\xi) -\nabla h_k(\xi)$
			\State $s \gets s_0/ \fnorm{\nabla \lt_{\calI_k}(\xi)}$. \label{algo:pg:step-size}
			\State \textbf{repeat} \label{algo:line-search-start}
			\State {~~~} $\xi^+ \gets \prox_{ s \rho} (\xi -  s \nabla \lt_{\calI_k}(\xi))$. \label{alg:get_beta+}
			\State {~~~} \textbf{break if} $\lt_{\calI_k}\left(\xi^+\right) 
						\leq  \lt_{\calI_k}(\xi) 
							+ 
							\ip{{\nabla \lt_{\calI_k} (\xi)} , \xi^+ - \xi }
							+\frac1{2 s} \fnorm{\xi^+ - \xi}^2$.\label{alg:pg_stopcon}
		\State {~~~} $s \gets \kappa s$. \label{algo:line-search-end}
		\State err $\gets d\left(\xi^+, \xi\right)$
		where  $d(x,y) := \frac {\fnorm{x-y}} {\max\{1,\fnorm{y}\}}$.
		\State $\xi \gets \xi^+$ and iter $ \gets $ iter $+1$.
		\EndWhile
		\State \textbf{end while}
		\State $\beta^{(t)}_{k,\pi} \gets \xi$.
	\end{algorithmic}
\end{algorithm}

\subsection{Tuning parameter selection and structure estimation}
Given an initial permutation $\pi_0$, we use BIC grid search to select tuning parameter $\lambda$ that is used in the group Lasso penalty~\eqref{eq:group-lasso} (line~\ref{alg:bic}, Algorithm~\ref{algo:general}). To construct the grid, we select 20 equal-space points of $\lambda^{(i)}$ in the log scale from the interval $[0.01,  0.1]$, where $\lambda = 0.1$ is sufficiently large for an empty graph in our test. We select the tuning parameter $\lambda^{(i)}$ that minimizes the BIC score, $\text{BIC} = 2 \ell_{[n]}(\betah^{(i)}) + (\log n) \calN(\betah^{(i)})$, where $\betah^{(i)} \in \calD(\pi_0)$ is the minimizer of $F(\beta)$ with penalty parameter $\lambda^{(i)}$, computed by Algorithm~\ref{algo:dist-iter-opt}, and $\calN(\betah^{(i)})$ is the number of free parameters in $\betah^{(i)}$.

An estimated GLDAG parameter $\betah$ is provided at the end of the DARLS annealing search, from which we can estimate a causal structure (line~\ref{alg:refine-beta}, Algorithm~\ref{algo:general}). Let $W$ be a $p \times p$ weighted adjacency matrix of a DAG such that $W_{ij} := \fnorm{\betah_{ij}}$. 
The use of a group Lasso regularizer helps to eliminate some edges in learning a DAG, but it may still result in false positive edges. Hence, we further refine estimated structures by setting $W_{ij}$ to zero if $|W_{ij}| < \alpha \max_{ij} |W_{ij}|$. One can adjust the value of $\alpha$ to achieve a desired sparsity level, especially when having prior knowledge. In our simulation tests, we fix $\alpha = 0.1$ to remove edges whose weights are relatively small compared to others.

\section{Theoretical guarantees}
\label{sec:theoretical} 

In this section, we study the convergence of the iterative distributed optimization algorithm (Algorithm~\ref{algo:dist-iter-opt}) and establish the consistency of the global minimizer of~\eqref{eq:learn-bn-obj}. As our method is primarily motivated by applications involving a large amount of distributed data, we develop theoretical results under the setting that $n$ is large and the number of variables $p$ stays fixed.

\subsection{Distributed estimate convergence}
\label{sec:convergence}

Recall the local iteration functions $\varphi_{k,\pi}$ defined in~\eqref{eq:local:update}. The overall iteration function for the distributed algorithm can be written as $\Phi_\pi(\cdot) := \sum_{k} \frac{n_k}{n} \varphi_{k,\pi}(\cdot)$ (line~\ref{algo:io-beta-update}, Algorithm~\ref{algo:dist-iter-opt}). Let $\Sigma := \ex[x_h x_h^\top]$ be the population second-moment matrix of the model. For a matrix $\beta$, let $\ball_F(\beta;r)$ denote the Frobenius ball of radius $r$ centered at $\beta$. We consider the case of numerical variables, i.e. $d_j=1$ for all $j$, which includes continuous and binary discrete random variables. The following theorem provides convergence guarantees on the distributed optimization algorithm represented by $\Phi_\pi$ for any fixed $\pi$. Let $\beh_\pi$ be any global minimizer of the global objective function, i.e.,
\begin{equation}\label{eq:global-minimizer}
    \beh_\pi\in\argmin_{\xi \in \calD(\pi)} \ell_{[n]}(\xi) + \rho(\xi),
\end{equation}
where $\rho(\xi)$ is a convex regularizer. Let $\Omega:=\bigcup_{\pi}\calD(\pi)  \subset \reals^{(p+1) \times p}$ be the parameter space of GLDAGs.  We recall that for $\theta \in \Omega$, $\theta_j$ denotes the $j$th column and that $\{x_h\}$ is an i.i.d. sample from a GLDAG model~\eqref{eq:expo-fam-def}.

\begin{theorem}\label{thm:conv:concrete}
Assume that the coordinates of $x_h$ are $T$-bounded, that is, $|x_h^j| \le T$ for all $ h \in [n]$ and $j \in [p]$. Let $\theta\in\Omega$ be any GLDAG parameter and $r > 0$, and set $R^*_1 = \max_j \norm{\theta_j}_1$ and $r_p := 2 r \sqrt p$. Let $\bl_p = \inf_{|t| \le T (r_p + R_1^*)} b''(t)$, and assume that  $b''(\cdot)$ is $\bu_p$-Lipschitz on $[-Tr_p, Tr_p]$. Define
\begin{align*}
	\zeta_n := \left(T^3 \frac{\psi\big(\bu_p (r+\frac{R^*_1}{\sqrt p})\big) +  b''(0)}{\bl_p  \lamin(\Sigma)}\right)	\frac{p^{3/2} \log (np) }{\sqrt{m}},	
\end{align*}
where $\psi(x) :=\max\{x,\sqrt x\}$ and $m := \min_{k} |\Ic_k|$. Assume further that $np \ge \max\{K+1,3\}$. There exist constants $c_1, C_1, C > 0$ such that if $C_1 T^2  \sqrt{p^2 \log (np)/ m} \le \lamin(\Sigma)$, then with probability at least $1- 3(np)^{-c_1} - \pr(\fnorm{\beh_\pi - {\theta}} > r)$,
\begin{align*}
	\fnorm{\Phi_\pi(\beta) - \widehat \beta_\pi} \le C \zeta_n\, \fnorm{\beta - \widehat \beta_\pi}, \quad \text{for all}\; \beta \in \ball_F(\widehat \beta_\pi, r).
\end{align*}
\end{theorem}

Theorem~\ref{thm:conv:concrete} applies to any $\theta\in\Omega$. It is natural to take $\theta$ to be $\beta_\pi^*$, the minimizer of the population loss defined as
\begin{equation}\label{eq:argmin-pop-loss}
    \beta_\pi^* := \argmin_{\xi \in \calD(\pi)}\ex[ \ell_{[n]}(\xi)].
\end{equation}
Since $\beh_\pi$ is a consistent estimate of ${\beta^*_\pi}$ for any $\pi$ (Theorem~\ref{prop:perm-const}, Section~\ref{sec:consistency}), that $\pr(\fnorm{\beh_\pi - \beta^*_\pi} > r)$ goes to zero as $n$ grows. Thus, with high probability,  the iteration operator $\Phi_\pi(\cdot)$ will be a contraction: the sequence $\{\beta^{(t)}_{\pi}\}_{t \ge 0}$ produced by the distributed algorithm converges geometrically to $\beh_\pi$ if  $C \zeta_n < 1$.  For fixed $p$, and for sufficiently large $r$ such that $\beta^{(0)}_{\pi}\in \ball_F(\beh_\pi,r)$, one can always satisfy the condition of $C \zeta_n < 1$ by taking $m$ (the minimum sample size per machine) large enough. Hence, Theorem~\ref{thm:conv:concrete} provides a quantitative lower bound on $m$ for the geometric convergence to kick in.

Theorem~\ref{thm:conv:concrete} is proved by establishing the uniform concentration of the Hessian of the GLDAG model~\eqref{eq:expo-fam-def} around its expectation over certain balls in the parameter space, and then invoking a general convergence result for the DANE algorithm which we derive in the Supplementary material (cf. Theorem~\ref{thm:dane:conv}). Note that establishing such uniform concentration in the GLDAG model is challenging due to the highly dependent and nonlinear relation among the coordinates  $\{x_h^j\}_{j=1}^p$. A technical tool in establishing the concentration of the Hessian is the Ledoux--Talagrand contraction theorem. In order to extend the argument to the multi-logit and generally vector-valued DAG models with $d_j > 1$, one needs a multivariate extension of the contraction theorem which is not available in literature at the moment. This extensions is, in principle, possible and we leave it for the future work. The $T$-boundedness assumption is trivially satisfied for binary and ordinal data, which are the primary focus of this work. Removing this assumption  requires another nontrivial extension of the Ledoux--Talagrand theorem or a completely new approach. Whether an unbounded version of Theorem~\ref{thm:conv:concrete} with similar guarantees holds is unclear to us at this point. 

\subsection{Consistency}
\label{sec:consistency}

Let us write $\psi(x; \beta) := -\log p(x \mid \beta)$ for the negative log-likelihood of a single sample $x$ from model~\eqref{eq:expo-fam-def}. We view $x$, $x_h$ and $\beta$ as vectors by concatenating the columns when dealing with $\psi(\cdot; \cdot)$, so that $\beta \in \reals^D$ for $D = d(d+1)$. The consistency results established in this section applies to the class of models in~\eqref{eq:expo-fam-def}, not restricting to numerical variables. 

Recall that {$F(\beta) = \ell_{[n]}(\beta) + \lambda_n\sum_{i,j}\fnorm{\beta_{ij}}$} is the global regularized {negative} log-likelihood and $\Omega$ is the GLDAG parameter space. The optimization problem~\eqref{eq:learn-bn-obj} is equivalent to $\min_{\beta \in \Omega} F(\beta)$. We denote by $\betah \in \Omega$ a global minimizer of $F(\beta)$ and $\beta^* \in \Omega$ the true parameter with the true DAG $\calG^*$. 
For any $\beta$, let us consider the (cross) Fisher information matrix
\begin{align*}
		I(\beta; \beta^*) := \ex_{\beta^*} \grad^2 \psi(x; \beta).
\end{align*}
We note that $\psi(x; \cdot)$ is a convex function for exponential families, and hence $I(\beta; \beta^*)$ is always positive semi-definite.
To establish the consistency of $\betah$ as well as that of $\betah_\pi$, used in Theorem~\ref{thm:conv:concrete}, for any fixed $\pi$, we make the following assumptions:

\begin{enumerate}[label=(A\arabic*),itemsep=0ex] 
    \item\label{ass:ident} The true DAG $\calG^*$ is identifiable.
    \item \label{ass:higher-order} For every $\pi$, there exists a neighborhood of $\beta^*_\pi$, denoted by $\text{nb}(\beta^*_\pi)$ and functions $M_{jkl}$ such that $\left|\f {\partial^3} {\partial \beta_j \partial\beta_k \partial\beta_l} \psi(x; \beta) \right| \leq M_{jkl}(x)$ for all $\beta \in \text{nb}(\beta^*_\pi)$, almost surely, and $\ex_{\beta^*}[ M_{jkl}(x)] <\infty$ for all $j,k$ and~$l$.
	\item \label{ass:fisher-and-others} For every $\pi$, we have
	\begin{align*}
		 \inf_{u\, \in \, D(\pi),\; \norm{u} = 1} \ip{u, I(\beta_\pi^*; \beta^*)u} > 0.
	\end{align*}
\end{enumerate} 
In~\ref{ass:higher-order}, it is impliclty assumed that $\psi(x;\cdot)$ is finite in $\text{nb}(\beta^*_\pi)$ almost surely.

Before stating our theoretical results, we define $\Pi^* := \{\pi:\; \beta_\pi^* = \beta^*\}$ which is exactly the set of permutations consistent with $\beta^*$, and in particular, it is nonempty. Below is a sketch of argument, and a full proof is provided in the supplement (Section~\ref{sec:const-proof}).
First, for any $\pi$ that is consistent with $\beta^*$, we have $\beta^* \in \mathcal D(\pi)$. A KL divergence argument then shows that $\beta^*$ is the unique solution of the optimization problem defining $\beta^*_\pi$. That is, any $\pi$ consistent with $\beta^*$ belongs to $\Pi^*$. Conversely, if $\pi \in \Pi^*$, then $\beta^* =  \beta_\pi^* \in \mathcal D(\pi)$, and hence $\pi$ is consistent with $\beta^*$. With this observation, we establish the desired consistency results in the following theorem.

\begin{theorem}\label{prop:perm-const}
Assume~\ref{ass:ident}--\ref{ass:fisher-and-others} and $\sqrt n \lambda_n = \calO_p(1)$. Then, 
\begin{enumerate}[(a)]
    \item For every $\pi$, $F(\cdot)$ has a unqiue minimizer $\betah_\pi$ over $\mathcal D(\pi)$ and
    \begin{align*}
        \sup_{\pi\in\calP}\fnorm{\betah_\pi - \beta_\pi^*} = \calO_p (n^{-1/2}).
    \end{align*}
    \item $F(\cdot)$ has a unique minimizer $\betah$ over~$\Omega$ (the space of DAGs) and 
    \begin{align*}
        \fnorm{\betah - \beta^*} = \calO_p (n^{-1/2}).
    \end{align*}
    \item With probability converging to one as $n \to \infty$, $\betah = \betah_{\widehat \pi}$ for some (sequence of) $\widehat \pi \in \Pi^*$. 
\end{enumerate}
\end{theorem}

Theorem~\ref{prop:perm-const} confirms that the Group Lasso regularized estimator $\betah$, defined as a global minimzier of $F$, is $\sqrt n$-consistent, and it will identify a correct topological sort $\widehat\pi\in\Pi^*$ in the large-sample limit. Moreover, the theorem also establishes the uniform consistency of restricted miminizers $\betah_{\pi}$ for all $\pi$, which is needed for Theorem~\ref{thm:conv:concrete}. Assumption~\ref{ass:ident} holds under mild conditions according to Proposition~\ref{prop:multi-node-ident2},
Assumption~\ref{ass:higher-order} is a standard regularity condition, and Assumption~\ref{ass:fisher-and-others} is related to the non-singularity of the second moment matrix $\ex_{\beta^*}(xx^\top)$. For example, consider the case $d_j = 1$ for all $j$ and assume that the elements of $x$ are $T$-bounded and let $R_{j} =  \max_{\pi} \norm{[\beta^*_\pi]_j}_1$, viewing $\beta^*_\pi$ as a matrix with $j$th column $[\beta^*_\pi]_j$. Then if $\inf_{|t|\le T R_j} b_j''(t) > 0$ for all $j$, non-singularity of $\ex_{\beta^*}(xx^\top)$ is sufficient for \ref{ass:fisher-and-others} to hold. 

\section{Results on simulated data}
\label{sec:numerical} \label{sec:results} 

Denote by $s_0$ the number of edges in a graph on $p$ nodes. We downloaded the following \texttt{networks} $(p, s_0)$ from the Bayesian networks repository~\citep{bnrep} to simulate data: \texttt{Asia} (8, 8), \texttt{Sachs} (11, 17), \texttt{Child} (20, 25), \texttt{Insurance} (27, 52), \texttt{Alarm} (37, 46), \texttt{Hailfinder} (56, 66) and \texttt{Hepar2} (70, 123). We generated data under the GLDAG model~\eqref{eq:expo-fam-def} (Section~\ref{sec:glm-data}) and other common DAG models (Section~\ref{sec:not-glm-data}), where the latter is to examine the robustness of our method against violation of its model assumptions.

\subsection{Methods}

We compared the DARLS algorithm to the following DAG structure learning methods: 
	the standard greedy hill climbing (HC) algorithm \citep{GMP11}, 
	the Peter-Clark (PC) algorithm \citep{pc_91}, 
	the max-min hill-climbing (MMHC) algorithm \citep{TEA06},
	the fast greedy equivalence search (FGES) \citep{C02, R17, QCK20},
	and the NOTEARS algorithm~\citep{ZA18-NOTEARS}.
Among these methods,  PC is a constraint-based method and MMHC is a hybrid method. The other three methods are score-based, where HC searches over DAGs, FGES searches over the equivalence classes, and NOTEARS uses continuous optimization to estimate DAG structure.

Following the practice for DAG learning on distributed data as in ~\cite{NY10, GJZ10, TWNA19}, we combined local estimates generated by a competing method to obtain a global graph estimate. Denote the dataset on a local machine by $D_k=\{x_h\}_{h \in \Ic_k}$, $k\in[K]$. We applied a competing method on each local dataset $D_k$ to obtain a completed partially directed acyclic graph (CPDAG) $A_k$, and then constructed a global graph using $\{A_k\}_{k=1}^K$. We used CPDAGs here because all the competing methods were developed under non-idenfitiable DAG models. Among the five competing methods, only PC and FGES output a CPDAG, and thus we converted estimated DAGs from the other methods to CPDAGs to obtain $A_k$. Given $\{A_k\}_1^K$, we counted occurrences of the three possible orientations between each pair $(i,j)$: $i\to j$, $i \leftarrow j$, or {$i-j$ (undirected)}. We then ranked orientations across all node pairs in the descending order of their counts, and sequentially added these edge orientations to an empty graph, as long as they would not introduce a directed cycle (a cycle consisting of all directed edges). By the end of this process, we had a partially directed graph. Lastly, we applied Meek's rules~\citep{M1995, C02} to maximally orient undirected edges, and hence constructed a global CPDAG estimate. 

\begin{remark}\label{remark:hc}
The HC global estimates had too many edges using this approach, because its local CPDAGs lacked consensus, resulting in a large number of candidate edges and much higher FP edges. To solve this problem, we did not add any edge between a node pair $(i,j)$ if the majority of the local graphs $\{A_k\}$ had no edges between them. In this way, global graphs estimated by HC became reasonably sparse compared to global estimates by the other methods.
\end{remark}

We implemented the DARLS algorithm in \MATLAB and used the following \texttt{packages} to run competing methods: \texttt{bnlearn}~\citep{S10} for the MMHC and HC algorithms, \texttt{pcalg}~\citep{Metc12} for the PC algorithms and \texttt{rcausal}~\citep{R17} for the FGES. The NOTEARS method was run with online Python code~\citep{notears_codes}. Competing methods were applied to each local dataset using a 2016 MacBook Pro (2.9 GHz Intel Core i5, 16 GB memory). Since DARLS is designed for distributed computing, it was run on a computer cluster at UCLA.

In this study, the DARLS algorithm (Algorithm~\ref{algo:general}) was initialized with a random permutation. According to the landscape of the objective function, we set the initial annealing temperature to $5 \cdot 10^{-2}$ and gradually decreased it to $5 \cdot 10^{-5}$ in a geometric fashion over a total of $10^3$ iterations. Note that since the log-likelihood has been normalized by the sample size, as in~\eqref{eq:glm-loss-beta}, the range of the objective function is quite small. For the PC algorithm, a significance level of $0.01$ was used to generate graphs with desired sparsity. FGES was applied with a significance level of $0.1$, which was the default value. For MMHC and HC methods, the maximum number of parents for a node was set to three. For NOTEARS, we used the default $\ell_2$ loss and the default threshold value. 

\subsection{{Accuracy metrics}}
\label{sec:numerical-accuracy}

Given estimates generated by the above methods, we use a few metrics to evaluate their structural accuracy. To standardize the performance metrics, we transform an estimated DAG into its CPDAG when the true DAG is not identifiable, before calculating the following metrics.

Let P, TP, FP, M, R be the number of estimated edges, true positive edges, false positive edges, missing edges and reversed edges, respectively. More specifically, P is the number of edges in the estimated graph, FP is the number of edges in the estimated graph skeleton but not in the true skeleton, and M counts the number of edges in the true skeleton but not in the skeleton of the estimated graph. TP reports the number of consistent edges between the estimated DAG/CPDAG and the true DAG/CPDAG, where a consistent edge must have the same orientation between the two nodes. There are two possible orientations for an edge in a DAG and three in a CPDAG. Lastly, the number of reversed edges $\text{R} = \text{P} - \text{TP} - \text{FP}$.  We then define structural Hamming distance, 
$\text{SHD} = \text{R} + \text{FP} + \text{M}$, as a combined metric. 
A method has higher structure learning accuracy if it achieves a lower SHD. 

\subsection{GLDAG data}
\label{sec:glm-data}

Logistic GLDAGs model~\eqref{eq:expo-fam-def} with $b_j(\beta_j^\top X) = \log(1+\exp(\beta_j^\top X))$ for all $j \in [p]$, was used to generate binary data $X_j\in\{0,1\}$, where the coefficient parameters $\{\beta_{ij}\}$ were uniformly sampled from $[-1.5, -0.8] \cup [0.8,1.5]$. We simulated 20 datasets for each network under two settings, $n=100p, K = 10$ and $n=10,000, K = 20$, where a total of $n$ observations were randomly assigned to $K$ local machines. Since GLDAGs are identifiable, we compare an estimated DAG by DARLS to the true DAG when calculating the SHD. 
For all other methods, we compare estimated and true CPDAGs since they do not assume an identifiable DAG.

Table~\ref{tab:shds-100p} reports the average performance metrics across 20 datasets for each of the seven graphs, using all six methods. Since NOTEARS estimates were too sparse to be competitive, using the default settings, we decreased the penalty tuning parameter to $10^{-4}$ from the suggested value of $10^{-1}$. Its results are listed only for two small graphs, \texttt{Asia} and \texttt{Sachs}. PC had difficulty generating estimates within a reasonable time limit (30 minutes per estimation) for the last two networks, \texttt{Hailfinder} and \texttt{Hepar2}, and hence it was removed from the comparisons on these two graphs. 
In both cases $n = 100p$ and $n = 10,000$, DARLS consistently achieved the lowest SHD among all methods for every network, demonstrating higher accuracy in estimating graphical structures. DARLS identified more TP edges than the other methods in almost every case. The refinement step via thresholding $\widehat{\beta}$ also helped to reduce SHD by cutting down FP edges. 

\begin{table}
\setlength{\tabcolsep}{3pt} 
\centering
\caption{DARLS against other methods on distributed logistic data.}
\label{tab:shds-100p}
\resizebox{\columnwidth}{!}{
\begin{tabular}{cc@{\hskip 0.1in}HHHHccccHHcH@{\hskip 0.1in}HHHHHccccHHcH}
\toprule
\centering
  {\texttt{Network}} & \multirow{2}{*}{Method} &  & &\multicolumn{9}{c}{$n=100p, K = 10$} 
    & & & & & & \multicolumn{9}{c}{$n=10,000, K = 20$}  \\
  $(p, s_0)$ &     &     n &   p &  s0 &    P &   TP &    R &   FP &    M &   TPR &   FPR &      SHD (sd) &      JI (sd) 
   &     &     n &   p &  s0 &    P &   TP &    R &   FP &    M &   TPR &   FPR &      SHD (sd) &      JI (sd) \\
\midrule
 \texttt{Asia} 
 & DARLS &   800 &   8 &   8 &   7.2 &   3.7 &   3.1 &   0.3 &   1.1 &  0.46 &  0.49 &   4.6 (1.6) &  0.33 (0.15) 
  & DARLS &  10000 &   8 &   8 &   7.5 &   4.5 &   3.0 &   0.0 &   0.5 &  0.57 &  0.39 &   3.5 (1.3) &  0.43 (0.17) \\
 (8, 8)  
      & MMHC &   800 &   8 &   8 &   7.0 &   2.6 &   3.0 &   1.4 &   2.4 &  0.33 &  0.60 &   6.7 (1.3) &  0.22 (0.06) 
      & MMHC &  10000 &   8 &   8 &   8.2 &   3.1 &   4.8 &   0.3 &   0.1 &  0.39 &  0.62 &   5.2 (1.0) &  0.25 (0.08) \\
      & FGES &   800 &   8 &   8 &   9.8 &   2.5 &   3.8 &   3.5 &   1.7 &  0.31 &  0.74 &   9.0 (2.4) &  0.17 (0.09) 
      & FGES &  10000 &   8 &   8 &  10.9 &   2.8 &   5.0 &   3.0 &   0.1 &  0.35 &  0.74 &   8.2 (2.2) &  0.19 (0.13) \\
      & HC &   800 &   8 &   8 &   1.7 &   0.8 &   0.8 &   0.0 &   6.3 &  0.11 &  0.45 &   7.2 (0.8) &  0.10 (0.10) 
      & HC &  10000 &   8 &   8 &   6.8 &   3.3 &   3.5 &   0.0 &   1.2 &  0.41 &  0.51 &   4.7 (0.7) &  0.29 (0.09) \\
      & PC &   800 &   8 &   8 &   6.2 &   2.1 &   3.1 &   0.8 &   2.7 &  0.27 &  0.63 &   6.7 (1.4) &  0.19 (0.10) 
      & PC &  10000 &   8 &   8 &   8.5 &   3.0 &   4.8 &   0.6 &   0.1 &  0.38 &  0.64 &   5.5 (1.5) &  0.24 (0.13) \\
      & NOTEARS &   800 &   8 &   8 &   5.7 &   0.5 &   0.9 &   4.2 &   6.6 &  0.06 &  0.92 &  11.8 (1.1) &  0.04 (0.05)  
      & NOTEARS &  10000 &   8 &   8 &   3.2 &   0.2 &   0.6 &   2.4 &   7.2 &  0.03 &  0.93 &  10.2 (1.1) &  0.02 (0.04) \vspace{2mm} \\
\rule{0pt}{2.5ex}    
\texttt{Sachs} 
& DARLS &  1100 &  11 &  17 &  15.1 &   7.5 &   7.2 &   0.5 &   2.4 &  0.44 &  0.50 &  10.0 (2.1) &  0.31 (0.10) 
      & DARLS &  10000 &  11 &  17 &  15.8 &   8.7 &   7.0 &   0.0 &   1.2 &  0.51 &  0.45 &   8.3 (2.3) &  0.37 (0.12) \\
(11, 17)        
      & MMHC &  1100 &  11 &  17 &  14.1 &   8.8 &   3.5 &   1.8 &   4.7 &  0.52 &  0.37 &  10.0 (4.7) &  0.44 (0.26) 
      & MMHC &  10000 &  11 &  17 &  16.8 &   6.8 &   9.2 &   0.8 &   1.1 &  0.40 &  0.60 &  11.1 (3.5) &  0.27 (0.18) \\
      & FGES &  1100 &  11 &  17 &  17.8 &   3.2 &   9.3 &   5.2 &   4.5 &  0.19 &  0.81 &  19.1 (3.1) &  0.11 (0.07) 
      & FGES &  10000 &  11 &  17 &  21.1 &   2.4 &  13.8 &   4.8 &   0.8 &  0.14 &  0.88 &  19.4 (3.5) &  0.07 (0.05) \\
      & HC &  1100 &  11 &  17 &   4.1 &   3.6 &   0.5 &   0.0 &  12.9 &  0.21 &  0.08 &  13.3 (1.4) &  0.21 (0.09) 
      & HC &  10000 &  11 &  17 &  10.8 &   6.8 &   4.0 &   0.0 &   6.2 &  0.40 &  0.35 &  10.2 (2.8) &  0.34 (0.17) \\
      & PC &  1100 &  11 &  17 &  12.6 &   4.7 &   6.5 &   1.4 &   5.8 &  0.28 &  0.60 &  13.7 (2.4) &  0.20 (0.12) 
      & PC &  10000 &  11 &  17 &  16.8 &   5.9 &  10.2 &   0.7 &   0.9 &  0.35 &  0.65 &  11.8 (4.4) &  0.25 (0.26) \\
      & NOTEARS &  1100 &  11 &  17 &   7.5 &   0.0 &   2.6 &   4.8 &  14.3 &  0.00 &  1.00 &  21.8 (1.4) &  0.00 (0.00) 
      & NOTEARS &  10000 &  11 &  17 &   5.8 &   0.0 &   2.3 &   3.5 &  14.7 &  0.00 &  1.00 &  20.6 (1.7) &  0.00 (0.00) \vspace{2mm}\\
\rule{0pt}{2.5ex}    
\texttt{Child} 
	& DARLS &  2000 &  20 &  25 &  23.0 &  15.1 &   7.4 &   0.5 &   2.5 &  0.60 &  0.34 &  10.4 (3.5) &  0.47 (0.14) 
	& DARLS &  10000 &  20 &  25 &  24.1 &  16.9 &   7.2 &   0.1 &   0.9 &  0.68 &  0.30 &   8.2 (3.7) &  0.54 (0.16) \\
(20, 25)        
      & MMHC &  2000 &  20 &  25 &  28.1 &   9.2 &  12.7 &   6.2 &   3.1 &  0.37 &  0.67 &  22.1 (3.7) &  0.21 (0.07) 
      & MMHC &  10000 &  20 &  25 &  29.1 &  10.2 &  14.6 &   4.3 &   0.3 &  0.41 &  0.65 &  19.2 (3.9) &  0.24 (0.09) \\
      & FGES &  2000 &  20 &  25 &  30.7 &   6.5 &  14.0 &  10.2 &   4.5 &  0.26 &  0.79 &  28.8 (4.7) &  0.13 (0.06) 
      & FGES &  10000 &  20 &  25 &  34.1 &  11.7 &  12.8 &   9.7 &   0.6 &  0.47 &  0.65 &  23.0 (5.0) &  0.25 (0.08)\\
      & HC &  2000 &  20 &  25 &  11.6 &   6.2 &   5.3 &   0.0 &  13.4 &  0.25 &  0.47 &  18.8 (2.8) &  0.21 (0.10) 
      & HC &  10000 &  20 &  25 &  20.1 &  13.3 &   6.8 &   0.0 &   4.9 &  0.53 &  0.35 &  11.7 (4.9) &  0.44 (0.20) \\
      & PC &  2000 &  20 &  25 &  25.6 &   6.5 &  14.4 &   4.7 &   4.0 &  0.26 &  0.74 &  23.1 (3.3) &  0.15 (0.05) 
      & PC &  10000 &  20 &  25 &  28.2 &   8.1 &  16.2 &   3.9 &   0.7 &  0.32 &  0.71 &  20.8 (4.5) &  0.19 (0.11)  \vspace{2mm}\\
\rule{0pt}{2.5ex}    
\texttt{Insurance} 
& DARLS &  2700 &  27 &  52 &  48.3 &  30.9 &  16.8 &   0.6 &   4.2 &  0.60 &  0.36 &  21.6 (5.0) &  0.45 (0.10) 
& DARLS &  10000 &  27 &  52 &  47.7 &  31.9 &  15.4 &   0.3 &   4.7 &  0.61 &  0.33 &  20.4 (4.3) &  0.47 (0.08) \\
(27, 52)        
      & MMHC &  2700 &  27 &  52 &  49.5 &  18.3 &  26.7 &   4.5 &   7.0 &  0.35 &  0.63 &  38.2 (3.7) &  0.22 (0.05) 
      & MMHC &  10000 &  27 &  52 &  53.2 &  22.1 &  27.1 &   4.1 &   2.9 &  0.42 &  0.58 &  34.0 (5.3) &  0.27 (0.07) \\
      & FGES &  2700 &  27 &  52 &  52.6 &  18.4 &  24.9 &   9.3 &   8.7 &  0.35 &  0.65 &  42.9 (5.1) &  0.22 (0.05) 
      & FGES &  10000 &  27 &  52 &  59.4 &  23.9 &  23.6 &  11.8 &   4.5 &  0.46 &  0.59 &  39.9 (7.0) &  0.28 (0.05) \\
      & HC &  2700 &  27 &  52 &  22.4 &  11.0 &  11.3 &   0.0 &  29.6 &  0.21 &  0.51 &  41.0 (4.1) &  0.18 (0.07) 
      & HC &  10000 &  27 &  52 &  32.3 &  20.4 &  11.9 &   0.0 &  19.7 &  0.39 &  0.37 &  31.6 (4.9) &  0.32 (0.09) \\
      & PC &  2700 &  27 &  52 &  46.2 &  15.8 &  26.9 &   3.5 &   9.2 &  0.30 &  0.66 &  39.6 (2.9) &  0.19 (0.03) 
      & PC &  10000 &  27 &  52 &  51.6 &  17.9 &  30.3 &   3.4 &   3.8 &  0.35 &  0.65 &  37.5 (4.4) &  0.21 (0.05) \vspace{2mm} \\
\rule{0pt}{2.5ex}    
\texttt{Alarm} 
     & DARLS &  3700 &  37 &  46 &  42.9 &  27.4 &  15.3 &   0.2 &   3.3 &  0.60 &  0.36 &  18.8 (2.3) &  0.45 (0.05) 
        & DARLS &  10000 &  37 &  46 &  43.4 &  27.1 &  16.2 &   0.1 &   2.8 &  0.59 &  0.37 &  19.1 (4.1) &  0.44 (0.09) \\
(37, 37)      
      & MMHC &  3700 &  37 &  46 &  53.9 &  17.9 &  25.8 &  10.3 &   2.4 &  0.39 &  0.67 &  38.5 (5.7) &  0.22 (0.05) 
      & MMHC &  10000 &  37 &  46 &  59.8 &  17.6 &  27.8 &  14.5 &   0.7 &  0.38 &  0.70 &  43.0 (5.7) &  0.20 (0.05) \\
      & FGES &  3700 &  37 &  46 &  51.6 &  19.2 &  22.1 &  10.3 &   4.7 &  0.42 &  0.63 &  37.1 (4.4) &  0.25 (0.05) 
      & FGES &  10000 &  37 &  46 &  63.6 &  23.5 &  21.2 &  18.9 &   1.3 &  0.51 &  0.63 &  41.4 (7.7) &  0.28 (0.06) \\
      & HC &  3700 &  37 &  46 &  29.7 &  10.9 &  18.8 &   0.0 &  16.3 &  0.24 &  0.64 &  35.0 (4.7) &  0.17 (0.08) 
      & HC &  10000 &  37 &  46 &  35.1 &  14.5 &  20.6 &   0.0 &  10.9 &  0.32 &  0.59 &  31.5 (4.9) &  0.22 (0.08) \\
      & PC &  3700 &  37 &  46 &  50.8 &  20.1 &  22.8 &   7.9 &   3.1 &  0.44 &  0.60 &  33.8 (3.2) &  0.26 (0.04) 
      & PC &  10000 &  37 &  46 &  58.5 &  18.9 &  26.4 &  13.3 &   0.8 &  0.41 &  0.68 &  40.5 (5.3) &  0.22 (0.06) \vspace{2mm}\\
\rule{0pt}{2.5ex}    
\texttt{Alarm} 
    & DARLS &  5600 &  56 &  66 &  63.0 &  39.2 &  23.0 &   0.8 &   3.8 &  0.59 &  0.38 &  27.5 (3.5) &  0.44 (0.05) 
	& DARLS &  10000 &  56 &  66 &   62.5 &  39.0 &  22.6 &   0.8 &   4.3 &  0.59 &  0.37 &  27.7 (4.6) &  0.44 (0.07) \\
(56, 66)
           & MMHC &  5600 &  56 &  66 &  80.7 &  16.4 &  44.6 &  19.6 &   5.0 &  0.25 &  0.80 &  69.2 (6.4) &  0.13 (0.02) 
           & MMHC &  10000 &  56 &  66 &  105.5 &  17.1 &  45.6 &  42.8 &   3.3 &  0.26 &  0.84 &  91.7 (8.7) &  0.11 (0.02) \\
           & FGES &  5600 &  56 &  66 &  75.0 &  23.0 &  39.0 &  13.0 &   4.0 &  0.35 &  0.69 &  56.0 (6.5) &  0.20 (0.04)            
           & FGES &  10000 &  56 &  66 &   90.8 &  21.7 &  41.0 &  28.1 &   3.3 &  0.33 &  0.76 &  72.5 (8.5) &  0.16 (0.03) \\
           & HC &  5600 &  56 &  66 &  53.0 &  14.8 &  38.2 &   0.0 &  13.0 &  0.22 &  0.72 &  51.2 (4.4) &  0.14 (0.05) 
           & HC &  10000 &  56 &  66 &   50.0 &  15.7 &  34.4 &   0.0 &  15.9 &  0.24 &  0.69 &  50.4 (6.6) &  0.16 (0.08) \vspace{2mm}\\
\rule{0pt}{2.5ex}    
\texttt{Hepar2} 
    & DARLS &  7000 &  70 &  123 &  112.3 &  72.9 &  37.9 &   1.6 &  12.2 &  0.59 &  0.35 &  51.7 (10.0) &  0.45 (0.07) 
	 & DARLS &  10000 &  70 &  123 &  112.6 &  75.5 &  35.8 &   1.3 &  11.7 &  0.61 &  0.33 &   48.8 (10.4) &  0.47 (0.08) \\
(70, 123)
       & MMHC &  7000 &  70 &  123 &  117.3 &  53.0 &  45.2 &  19.1 &  24.7 &  0.43 &  0.55 &   89.0 (7.6) &  0.28 (0.04) 
       & MMHC &  10000 &  70 &  123 &  153.4 &  51.0 &  55.9 &  46.5 &  16.1 &  0.41 &  0.67 &  118.5 (10.1) &  0.23 (0.03) \\
       & FGES &  7000 &  70 &  123 &  131.5 &  75.8 &  36.8 &  18.9 &  10.4 &  0.62 &  0.42 &   66.2 (9.9) &  0.43 (0.05)        
       & FGES &  10000 &  70 &  123 &  150.6 &  76.4 &  34.3 &  39.9 &  12.3 &  0.62 &  0.49 &   86.5 (11.9) &  0.39 (0.04) \\
       & HC &  7000 &  70 &  123 &   90.2 &  60.2 &  29.9 &   0.0 &  32.9 &  0.49 &  0.33 &   62.8 (5.9) &  0.39 (0.04) 
       & HC &  10000 &  70 &  123 &   79.5 &  49.5 &  30.1 &   0.0 &  43.5 &  0.40 &  0.38 &   73.5 (11.0) &  0.33 (0.09) \\
    \bottomrule
\end{tabular}
}
\end{table}

\begin{figure}[t]
	\centering
	\includegraphics[width=0.8\textwidth]{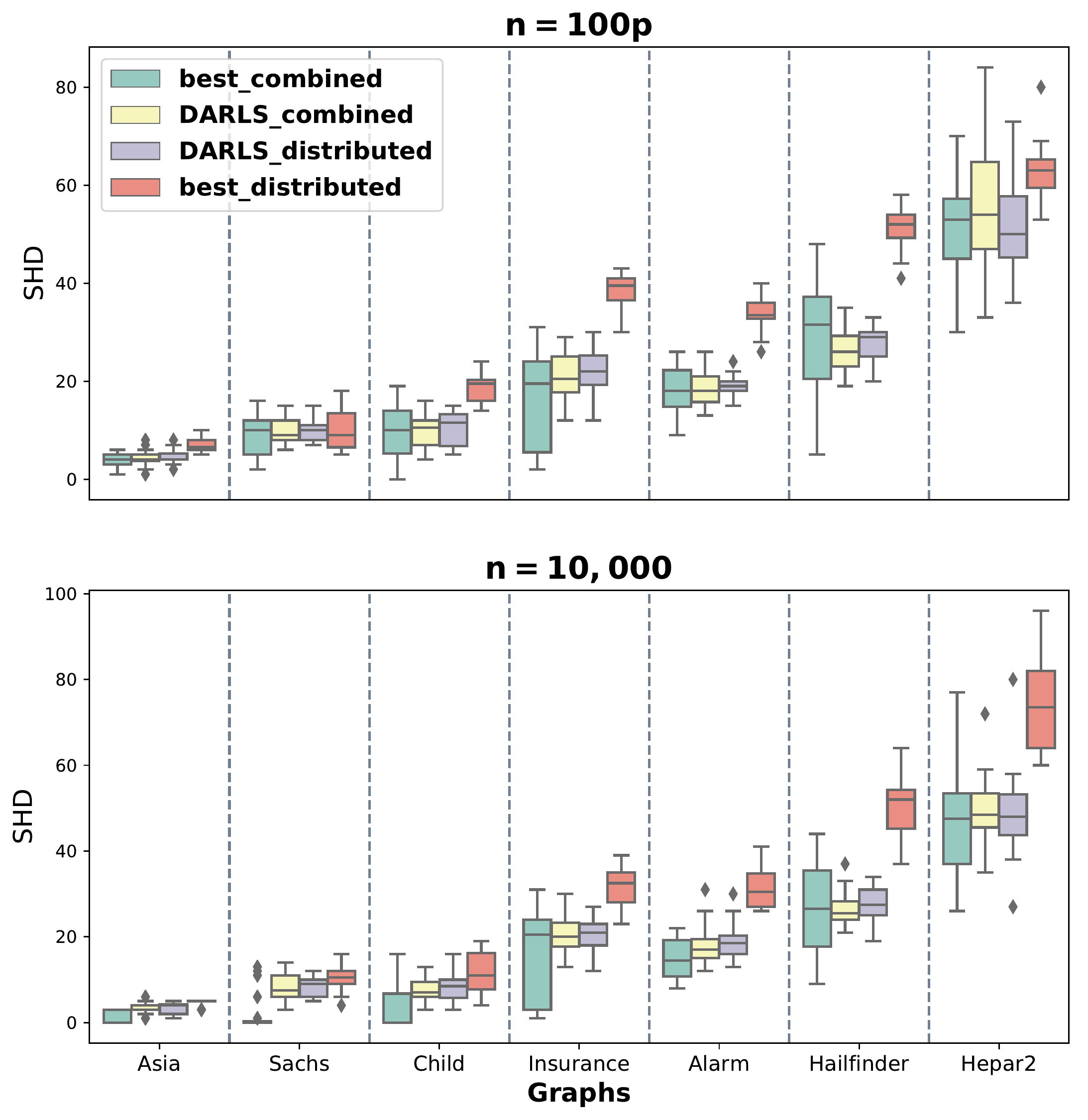}
        \caption{SHD comparison between DARLS and the best competing method using combined or distributed data. For each graph, the box plots from left to right report the results for the best competing method on combined data (best-combined), DARLS on combined combined, DARLS on distributed data and the best competing method on distributed data (best-distributed).} 
        \label{fig:shds}
\end{figure}

To examine the loss of accuracy in estimating network structures on distributed data, we applied each method to  combined data by pooling the $K$ local datasets. For brevity in reporting the results, we chose the best performance on each network among the five competing methods, called the best competing method, to compare with DARLS. Figure~\ref{fig:shds} shows the performance, in terms of SHDs, of DARLS and the best competing method on distributed and combined data. First, we observe that, applied to either distributed or combined data, DARLS achieves similar SHD values. The SHD difference between the combined and distributed data of DARLS is much smaller than that of the best competing method, demonstrating that DARLS is more effective in using distributed data. Second, consistent with Table~\ref{tab:shds-100p}, DARLS always outperformed the best competing method significantly when applied to distributed data (best-distributed). Furthermore, DARLS on distributed data shows a substantial overlap in the distribution of SHD, with the best competing method on combined data (best-combined), which is either HC or FGES for $n=100p$ and HC, MMHC or FGES for $n = 10,000$. Such overlaps indicate the highly competitive performance of our distributed learning algorithm. The variability in SHD of DARLS is in general smaller than that of the best competing method, showing a higher consistency across different datasets.

\begin{figure}
	\centering
	\begin{subfigure}[b]{0.485\textwidth}
         	\centering
         	\includegraphics[width=\textwidth]{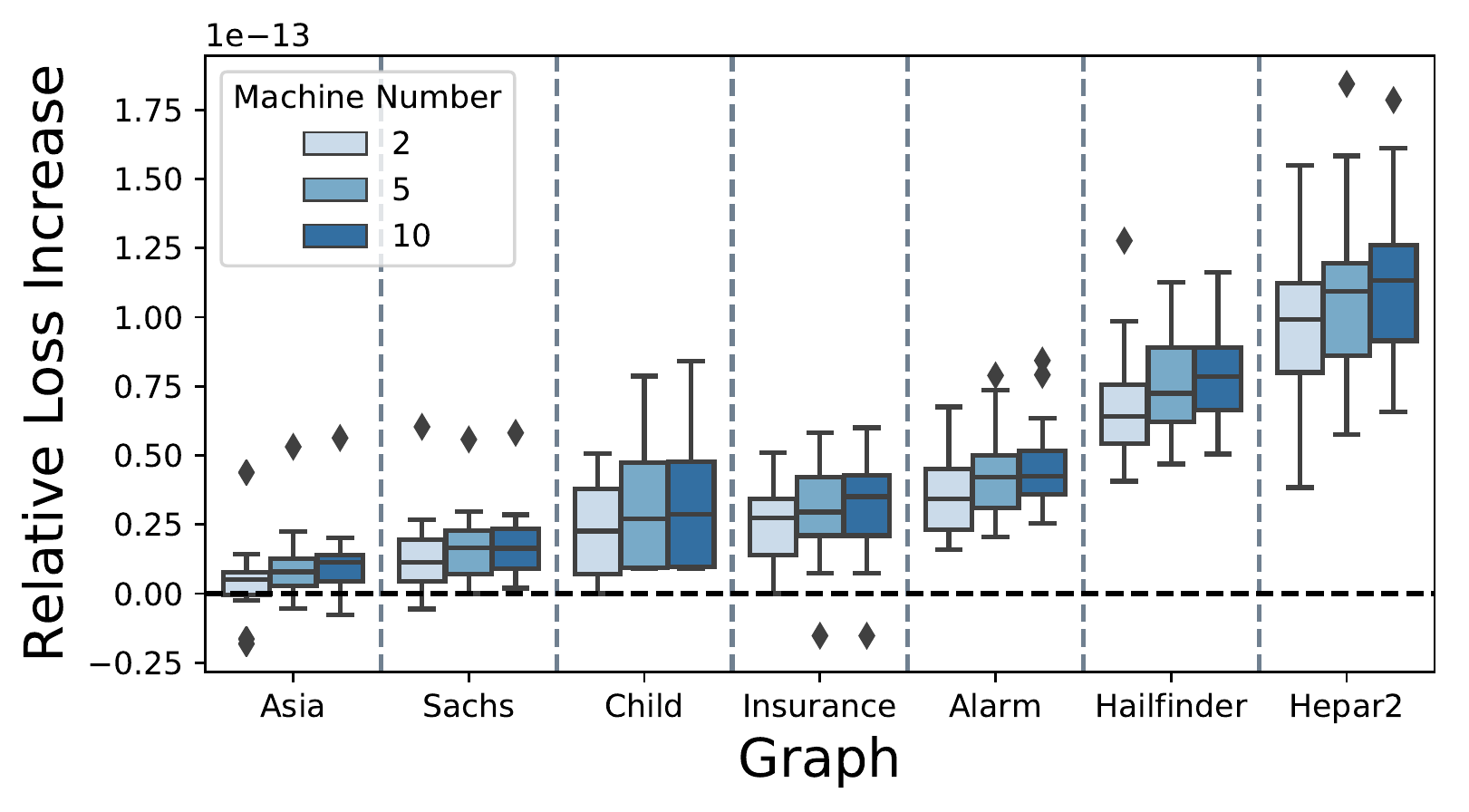}
         	\caption{Relative loss distribution.} 
         	\label{fig:fval}
	\end{subfigure}
       \hfill
       \begin{subfigure}[b]{0.485\textwidth}
         	\centering
         	\includegraphics[width=\textwidth]{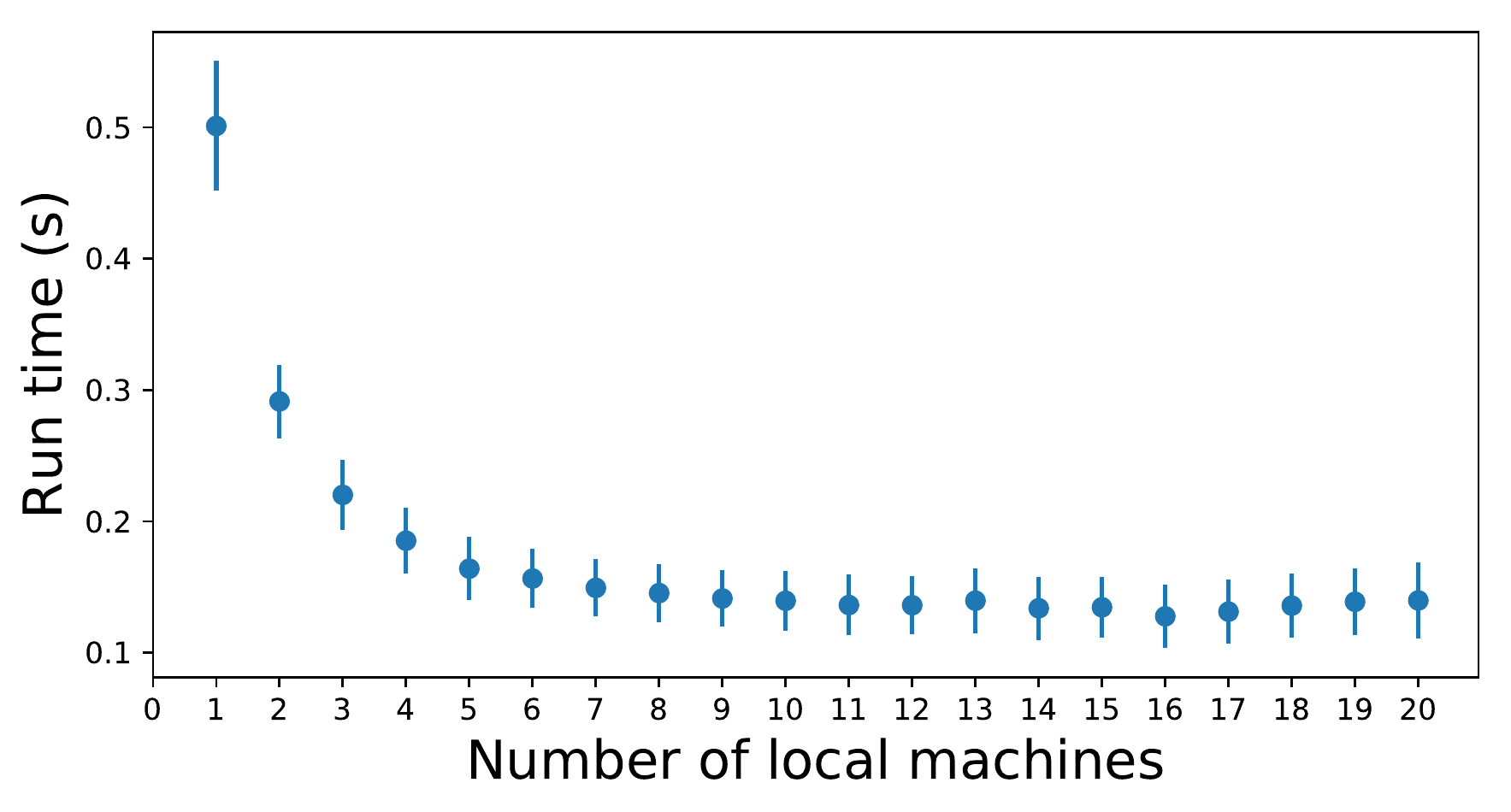}
         	\caption{Computation time distribution.}
         	\label{fig:runtime}
	\end{subfigure}
       \caption{Accuracy and computation time comparison for $f(\pi)$.
       computation time comparison, with mean and standard deviations plotted, is performed over 20 datasets generated from \texttt{Insurance}.}
        \label{fig:fval_rt}
\end{figure}

To quantify the accuracy of distributed optimization using Algorithm~\ref{algo:dist-iter-opt}, we computed permutation scores $f(\pi)$~\eqref{eq:learn-bn-obj} under various values of $K \in \{1, 2, 5, 10\}$ for a fixed $\pi$. For each value of $K$, we fixed the tuning parameter $\lambda$, permutation $\pi$ and all internal computation parameters to ensure the only $K$ varied in the calculation of $f(\pi)$. Let $f^{(K)}$ be the value of $f(\pi)$ computed by $K$ local machines, and $\Delta f^{(K)} := f^{(K)} - f^{(1)}$ be the relative increase in the loss $f^{(K)}$. Figure~\ref{fig:fval} shows $\{\Delta f^{(K)}: K = 2, 5, 10\}$  across all the networks. 
The values of $\Delta f^{(K)}$ is in the order of $10^{-13}$, verifying $f(\pi)$ is essentially identical using either the overall ($K = 1$) or distributed data ($K \geq 2$). Since the number of iterations is fixed, $\Delta f^{(K)}$ increases with the network size. 

We also examined computation time of finding $f^{(K)}$ for $K \in [20]$ using the 20 \texttt{Insurance} datasets. 
In each test, the same data were split and distributed to different number $K$ of local machines to solve the optimization problem~\eqref{eq:learn-bn-obj}, with all other parameters fixed. Figure~\ref{fig:runtime} shows the computation time versus $K$. As expected when distributing a complicated task, computing $f(\pi)$ requires less time if using more machines. However, the reduction in computation time reaches approximately a stable level after $K=10$, indicating a trade-off between the gains from parallel computation and the communication overhead. Additionally, the smallest local sample size $m$ decreases when $K$ is large, and Theorem~\ref{thm:conv:concrete} shows that would slow down the convergence of Algorithm~\ref{algo:dist-iter-opt}.

\subsection{{Other data generation models}}
\label{sec:not-glm-data}

To test the robustness of DARLS against violations of its model assumptions, we also generated data from different DAG models. Particularly, we used threshold-Gaussian and multinomial DAGs to generate discrete data, and then compared DARLS with the other methods.

For the threshold-Gaussian DAG model, we first generated continuous variables $Y_1,\ldots,Y_p$ using Gaussian structural equation models $Y_j = B_j^\top \text{PA}_j + \epsilon_j$ where $\epsilon_j$ is a Gaussian noise from $\calN(0, 1)$ and each coefficient parameter in $B_j\in \mathbb{R}^{s_j}$ was sampled uniformly from $[-1.2, -0.8] ~\bigcup ~[0.8, 1.2]$, where $s_j=|\text{PA}_j|$. Then, we thresheld these continuous values to generate binary data $X_j=\calI(Y_j > c_j)$, where $c_j$ is the sample mean of $Y_j$. We used two-component mixture Gaussian to simulate continuous data for root variables in $Y_j$, each drawn from $\calN(1, 1)$ and $\calN(-1, 1)$ with an equal probability. Under this design, most of the $Y_j$'s were bi-modal in distribution, which greatly increased the robustness in thresholding. Such conversion between continuous and categorical variables has been used for modeling BNs in prior work~\citep{L2010, C2020, Setc2021}, but the threshold-Gaussian DAG model is not the underlying model for any of six methods in our comparison.

We also simulated multinomial data from contingency tables provided in the BN repository~\citep{bnrep}. We made a few modifications to the contingency tables to ensure that
	(1) the number of states per variable was at most three, and 
	(2) the marginal probability of any state was at least $0.1$ for every variable by merging states.
Due to the high structure complexity of \texttt{Hailfinder} and \texttt{Hepar2}, the original contingency tables of a few nodes were used without modifications, resulting in marginal probability less than $0.1$ for some states. We note that the multinomial DAG model is the underlying model for the majority of the competing methods, including HC, PC, FGES and MMHC. Therefore, comparisons on these data would also test if the GLDAG model can approximate the multinomial model reasonably well.

\begin{table}
\setlength{\tabcolsep}{4pt} 
\centering
\caption{DARLS against other methods when its model assumption is violated.}
\label{tab:shds-gaussian-multinomial}
\resizebox{\columnwidth}{!}{
\begin{tabular}{cc@{\hskip 0.1in}HHHHccccHHcH@{\hskip 0.1in}HHHHHccccHHcH}
\toprule
\centering
  {Network} & \multirow{2}{*}{Method} &  & &\multicolumn{9}{c}{threshold-Gaussian} 
    & & & & & & \multicolumn{9}{c}{Multinomial}  \\
  $(p, s_0)$ &     &     n &   p &  s0 &    P &   TP &    R &   FP &    M &   TPR &   FPR &      SHD (sd) &      JI (sd) 
   &     &     n &   p &  s0 &    P &   TP &    R &   FP &    M &   TPR &   FPR &      SHD (sd) &      JI (sd) \\
\midrule
\texttt{Asia} 
        & DARLS &  10000 &   8 &    8 &    8.7 &   6.2 &   1.6 &    0.9 &    0.3 &  0.77 &  0.26 &    2.8 (3.5) &  0.69 (0.36) 
      & DARLS &  10000 &   8 &   8 &   7.1 &   7.0 &   0.0 &   0.1 &   1.0 &  0.88 &  0.01 &   1.1 (0.4) &  0.87 (0.04) \\
(8,8)      
       & MMHC &  10000 &   8 &    8 &    4.0 &   1.1 &   0.7 &    2.2 &    6.2 &  0.14 &  0.75 &    9.1 (1.7) &  0.11 (0.13) 
      & MMHC &  10000 &   8 &   8 &   6.5 &   4.8 &   1.2 &   0.5 &   2.0 &  0.59 &  0.25 &   3.8 (1.3) &  0.50 (0.10) \\
       & FGES &  10000 &   8 &    8 &    4.5 &   0.2 &   0.8 &    3.4 &    6.9 &  0.03 &  0.93 &   11.1 (1.5) &  0.02 (0.04) 
      & FGES &  10000 &   8 &   8 &  13.1 &   5.8 &   2.2 &   5.0 &   0.0 &  0.72 &  0.55 &   7.3 (2.0) &  0.39 (0.12) \\
       & HC &  10000 &   8 &    8 &    8.8 &   4.8 &   2.7 &    1.4 &    0.5 &  0.60 &  0.45 &    4.5 (3.3) &  0.47 (0.33) 
      & HC &  10000 &   8 &   8 &   8.1 &   7.8 &   0.1 &   0.1 &   0.0 &  0.98 &  0.02 &   0.2 (0.9) &  0.97 (0.13) \\
       & PC &  10000 &   8 &    8 &   14.2 &   2.2 &   5.5 &    6.6 &    0.3 &  0.28 &  0.84 &   12.4 (2.1) &  0.12 (0.07) 
      & PC &  10000 &   8 &   8 &   6.3 &   2.5 &   3.5 &   0.3 &   2.0 &  0.31 &  0.60 &   5.8 (0.7) &  0.21 (0.05) \\
       & NOTEARS &  10000 &   8 &    8 &    6.0 &   0.6 &   1.3 &    4.2 &    6.1 &  0.07 &  0.90 &   11.6 (2.0) &  0.05 (0.07)
      & NOTEARS &  10000 &   8 &   8 &   5.3 &   0.6 &   0.9 &   3.8 &   6.5 &  0.07 &  0.89 &  11.2 (1.7) &  0.05 (0.07) \vspace{2mm}\\
\rule{0pt}{2.5ex}    
\texttt{Sachs} 
        & DARLS &  10000 &  11 &   17 &   11.3 &   7.0 &   2.9 &    1.5 &    7.2 &  0.41 &  0.37 &   11.5 (3.0) &  0.35 (0.17) 
        & DARLS &  10000 &  11 &  17 &  14.1 &  10.9 &   3.1 &   0.1 &   3.0 &  0.64 &  0.23 &   6.2 (3.3) &  0.57 (0.24) \\
(11, 17)       
       & MMHC &  10000 &  11 &   17 &    5.1 &   2.0 &   1.6 &    1.6 &   13.4 &  0.12 &  0.59 &   16.6 (2.5) &  0.10 (0.13) 
      & MMHC &  10000 &  11 &  17 &  14.3 &   3.1 &  11.2 &   0.1 &   2.8 &  0.18 &  0.78 &  14.0 (1.1) &  0.11 (0.04) \\
       & FGES &  10000 &  11 &   17 &    6.3 &   1.6 &   2.4 &    2.4 &   13.1 &  0.09 &  0.70 &   17.8 (1.9) &  0.08 (0.05) 
      & FGES &  10000 &  11 &  17 &  19.1 &   4.4 &  11.8 &   2.9 &   0.8 &  0.26 &  0.76 &  15.5 (4.5) &  0.16 (0.19) \\
       & HC &  10000 &  11 &   17 &   13.5 &   6.2 &   5.2 &    2.1 &    5.7 &  0.36 &  0.54 &   12.9 (2.7) &  0.26 (0.12) 
      & HC &  10000 &  11 &  17 &  12.6 &   6.7 &   5.9 &   0.0 &   4.5 &  0.39 &  0.48 &  10.3 (4.3) &  0.33 (0.25) \\
       & PC &  10000 &  11 &   17 &   17.2 &   4.0 &   8.3 &    4.9 &    4.7 &  0.24 &  0.77 &   17.9 (2.7) &  0.14 (0.08) 
      & PC &  10000 &  11 &  17 &  10.1 &   8.8 &   1.1 &   0.1 &   7.0 &  0.52 &  0.13 &   8.3 (2.5) &  0.49 (0.17) \\
       & NOTEARS &  10000 &  11 &   17 &    8.7 &   0.0 &   2.6 &    6.0 &   14.4 &  0.00 &  1.00 &   23.1 (1.3) &  0.00 (0.00)
      & NOTEARS &  10000 &  11 &  17 &   8.6 &   0.1 &   2.8 &   5.7 &  14.2 &  0.00 &  1.00 &  22.6 (2.4) &  0.00 (0.01) \vspace{2mm}\\
\rule{0pt}{2.5ex}    
\texttt{Child} 
& DARLS &  10000 &  20 &   25 &   23.1 &  10.1 &   9.3 &    3.7 &    5.6 &  0.40 &  0.56 &   18.6 (5.7) &  0.29 (0.17) 
& DARLS &  10000 &  20 &  25 &  22.7 &  18.1 &   4.0 &   0.6 &   2.9 &  0.72 &  0.21 &   7.5 (4.3) &  0.63 (0.20) \\
(20, 25)
       & MMHC &  10000 &  20 &   25 &    9.6 &   3.0 &   4.2 &    2.5 &   17.9 &  0.12 &  0.67 &   24.5 (2.5) &  0.10 (0.07) 
      & MMHC &  10000 &  20 &  25 &  20.9 &  12.2 &   8.2 &   0.5 &   4.7 &  0.49 &  0.42 &  13.3 (3.9) &  0.38 (0.14) \\
       & FGES &  10000 &  20 &   25 &   12.2 &   1.6 &   2.8 &    7.9 &   20.6 &  0.06 &  0.88 &   31.4 (2.1) &  0.04 (0.04) 
      & FGES &  10000 &  20 &  25 &  28.2 &  14.4 &   9.9 &   3.9 &   0.6 &  0.58 &  0.49 &  14.4 (2.9) &  0.38 (0.08)  \\
       & HC &  10000 &  20 &   25 &   23.1 &   7.2 &  13.6 &    2.4 &    4.3 &  0.29 &  0.69 &   20.2 (3.4) &  0.18 (0.08) 
      & HC &  10000 &  20 &  25 &  20.1 &  14.9 &   5.0 &   0.1 &   5.0 &  0.60 &  0.26 &  10.2 (2.3) &  0.50 (0.10) \\
       & PC &  10000 &  20 &   25 &   53.2 &   7.8 &  13.6 &   31.8 &    3.5 &  0.31 &  0.85 &   49.0 (4.4) &  0.11 (0.04) 
      & PC &  10000 &  20 &  25 &  17.4 &   6.0 &  10.8 &   0.6 &   8.2 &  0.24 &  0.66 &  19.6 (2.5) &  0.17 (0.07) \vspace{2mm}\\
\rule{0pt}{2.5ex}    
\texttt{Insurance}
        & DARLS &  10000 &  27 &   52 &   38.8 &  11.8 &  13.2 &   13.7 &   26.9 &  0.23 &  0.69 &   53.8 (8.2) &  0.15 (0.08) 
        & DARLS &  10000 &  27 &  52 &  35.2 &  15.7 &  16.1 &   3.5 &  20.2 &  0.30 &  0.55 &  39.8 (7.6) &  0.23 (0.09) \\
(27, 52)       
       & MMHC &  10000 &  27 &   52 &   21.2 &   4.3 &   7.3 &    9.6 &   40.4 &  0.08 &  0.78 &   57.2 (5.8) &  0.06 (0.04) 
      & MMHC &  10000 &  27 &  52 &  32.8 &  16.6 &  15.2 &   0.9 &  20.2 &  0.32 &  0.49 &  36.3 (2.9) &  0.25 (0.05) \\
       & FGES &  10000 &  27 &   52 &   26.0 &   3.0 &   6.8 &   16.1 &   42.1 &  0.06 &  0.88 &   65.1 (3.9) &  0.04 (0.02)  
      & FGES &  10000 &  27 &  52 &  60.0 &  23.1 &  16.6 &  20.4 &  12.3 &  0.44 &  0.61 &  49.2 (6.0) &  0.26 (0.04) \\
       & HC &  10000 &  27 &   52 &   36.9 &   8.9 &  14.8 &   13.1 &   28.2 &  0.17 &  0.76 &   56.1 (4.7) &  0.11 (0.04) 
      & HC &  10000 &  27 &  52 &  33.9 &  22.5 &  10.7 &   0.8 &  18.9 &  0.43 &  0.34 &  30.2 (4.9) &  0.36 (0.09) \\
       & PC &  10000 &  27 &   52 &   69.8 &   9.3 &  20.2 &   40.2 &   22.4 &  0.18 &  0.87 &   82.8 (6.6) &  0.08 (0.02) 
      & PC &  10000 &  27 &  52 &  28.9 &   7.3 &  20.7 &   0.9 &  24.0 &  0.14 &  0.75 &  45.6 (2.8) &  0.10 (0.04)  \vspace{2mm} \\
\rule{0pt}{2.5ex}    
\texttt{Alarm}
	& DARLS &  10000 &  37 &   46 &   37.6 &  14.4 &  14.9 &    8.2 &   16.6 &  0.31 &  0.62 &   39.8 (7.4) &  0.21 (0.10) 
    & DARLS &  10000 &  37 &  46 &  42.0 &  18.1 &  17.6 &   6.2 &  10.2 &  0.39 &  0.57 &  34.0 (4.0) &  0.26 (0.06) \\
(37, 46)
       & MMHC &  10000 &  37 &   46 &   18.2 &   5.2 &   4.8 &    8.2 &   36.0 &  0.11 &  0.71 &   48.9 (3.9) &  0.09 (0.03) 
      & MMHC &  10000 &  37 &  46 &  40.9 &  24.6 &  13.9 &   2.3 &   7.4 &  0.54 &  0.40 &  23.6 (2.6) &  0.40 (0.05) \\
       & FGES &  10000 &  37 &   46 &   20.5 &   2.6 &   1.8 &   16.1 &   41.6 &  0.06 &  0.88 &   59.5 (4.7) &  0.04 (0.04) 
      & FGES &  10000 &  37 &  46 &  66.7 &  36.6 &   4.4 &  25.6 &   5.0 &  0.80 &  0.45 &  35.0 (6.0) &  0.49 (0.06) \\  
       & HC &  10000 &  37 &   46 &   45.5 &  18.9 &  16.9 &    9.8 &   10.2 &  0.41 &  0.59 &   36.9 (6.1) &  0.26 (0.09) 
      & HC &  10000 &  37 &  46 &  41.8 &  27.1 &  12.3 &   2.4 &   6.5 &  0.59 &  0.35 &  21.2 (4.5) &  0.45 (0.10) \\
       & PC &  10000 &  37 &   46 &   75.2 &  10.2 &  25.6 &   39.5 &   10.2 &  0.22 &  0.86 &   75.2 (4.8) &  0.09 (0.02) 
      & PC &  10000 &  37 &  46 &  38.6 &  11.6 &  25.2 &   1.8 &   9.2 &  0.25 &  0.70 &  36.2 (2.4) &  0.16 (0.03) \vspace{2mm}\\
\rule{0pt}{2.5ex}    
\texttt{Hailfinder}
	    & DARLS &  10000 &  56 &   66 &   45.4 &  16.5 &  17.1 &   11.8 &   32.4 &  0.25 &  0.64 &   61.3 (7.9) &  0.18 (0.07) 
        & DARLS &  10000 &  56 &  66 &  35.2 &  16.6 &  15.2 &   3.5 &  34.2 &  0.25 &  0.53 &  52.9 (5.5) &  0.20 (0.06)\\
(56, 66)
        & MMHC &  10000 &  56 &   66 &   28.4 &  15.1 &   4.8 &    8.4 &   46.1 &  0.23 &  0.46 &   59.4 (4.3) &  0.19 (0.04) 
        & MMHC &  10000 &  56 &  66 &  41.5 &  14.3 &  22.4 &   4.8 &  29.3 &  0.22 &  0.65 &  56.4 (4.1) &  0.16 (0.03) \\
        & FGES &  10000 &  56 &   66 &   20.6 &   2.2 &   1.8 &   16.6 &   62.0 &  0.03 &  0.90 &   80.3 (3.4) &  0.03 (0.02) 
        & FGES &  10000 &  56 &  66 &  69.7 &  28.1 &  23.9 &  17.6 &  14.0 &  0.43 &  0.60 &  55.6 (6.3) &  0.26 (0.07) \\
        & HC &  10000 &  56 &   66 &   62.8 &  32.9 &  20.4 &    9.4 &   12.7 &  0.50 &  0.47 &   42.5 (9.2) &  0.35 (0.12) 
        & HC &  10000 &  56 &  66 &  43.5 &  23.8 &  18.0 &   1.7 &  24.2 &  0.36 &  0.45 &  44.0 (2.3) &  0.28 (0.03) \vspace{2mm} \\
\rule{0pt}{2.5ex}    
\texttt{Hepar2}
	& DARLS &  10000 &  70 &  123 &   66.5 &  14.8 &  24.6 &   27.1 &   83.7 &  0.12 &  0.78 &  135.4 (8.6) &  0.08 (0.03) 
       & DARLS &  10000 &  70 &  123 &   49.9 &  18.1 &  28.6 &   3.2 &  76.3 &  0.15 &  0.64 &   108.1 (4.8) &  0.12 (0.04) \\
(70, 123)
       & MMHC &  10000 &  70 &  123 &   42.9 &   4.5 &  12.2 &   26.2 &  106.4 &  0.04 &  0.90 &  144.8 (3.6) &  0.03 (0.01) 
       & MMHC &  10000 &  70 &  123 &  111.7 &  35.0 &  25.1 &  51.5 &  62.9 &  0.28 &  0.69 &   139.4 (8.7) &  0.18 (0.03)\\
       & FGES &  10000 &  70 &  123 &   59.5 &   9.3 &   6.8 &   43.4 &  106.8 &  0.08 &  0.85 &  157.0 (6.9) &  0.05 (0.02) 
       & FGES &  10000 &  70 &  123 &  151.3 &  56.8 &  20.0 &  74.5 &  46.2 &  0.46 &  0.62 &  140.7 (11.0) &  0.26 (0.02) \\  
       & HC &  10000 &  70 &  123 &   82.2 &  29.8 &  27.4 &   25.0 &   65.8 &  0.24 &  0.64 &  118.2 (9.4) &  0.17 (0.06) 
       & HC &  10000 &  70 &  123 &   49.8 &  28.1 &  19.8 &   1.9 &  75.1 &  0.23 &  0.43 &    96.8 (7.8) &  0.20 (0.06) \\
\bottomrule
\end{tabular}
}
\end{table}

Table~\ref{tab:shds-gaussian-multinomial} reports the structural estimation accuracy of each method on data generated by the threshold-Gaussian and the mulitnomial DAG models, with $n=10,000$ and $K=20$. When the underlying model is a threshold-Gaussian DAG, DARLS achieved the lowest SHD for 4 networks, i.e. \texttt{Asia}, \texttt{Sachs}, \texttt{Child} and \texttt{Insurance}. For data simulated from multinomial models, DARLS still could estimate the network with the lowest SHD for \texttt{Sachs} and \texttt{Child}. For most of the other cases, DARLS was only worse than HC, but better than or at least comparable to the other methods. Note that we specifically tuned the way for HC to combine local estimates and build a global graph, due to its lack of consensus among local estimates (Remark~\ref{remark:hc}). It is encouraging to see that DARLS uniformly outperformed FGES, a consistent score-based method under the multinomial DAG model~\citep{C02}, highlighting the effectiveness of DARLS in using distributed data. This also suggests that the GLDAG~\eqref{eq:expo-fam-def} can be a pretty good approximation to the commonly used multinomial DAG model for discrete data. This study confirms that DARLS indeed performs relatively well on data generated from different DAG models, which is important for its practical use. This is further demonstrated by the application to a real dataset in next section.

\section{Real data application} 
\label{sec:real-data}

In this section, we apply our method to the ChIP-Seq data generated by \cite{Chen_etc_cell_08}. The dataset contains the DNA binding sites of 12 transcription factors (TFs) in mouse embryonic stem cells: {\em Smad1, Stat3, Sox2, Pou5f1, Nanog, Esrrb, Tcfcp2l1, Klf4, Zfx, E2f1, Myc,} and {\em Mycn}. For each TF, an association strength score, which is the weighted sum of the corresponding ChIP-Seq signal strength, was calculated for each of the 18,936 genes \citep{OZW09}. Roughly speaking, this score can be understood as a measure of the binding strength of a TF to a gene. Following the same preprocessing in \cite{WZ21}, the genes with zero association scores were removed from our analysis. Accordingly, our observed data matrix, of size $n \times p = 8462 \times 12$, contains the association scores of 12 TFs over 8,462 genes. We aim to build a causal network that reveals how these 12 TFs affect each other's binding to genes. The associate scores of a TF are typically bimodal, and they were discretized before network estimation; see Figure~\ref{fig:chip-dis-example} in the supplemental material for illustration of the discretization.

We distributed this dataset across $K=20$ local machines and applied DARLS, HC, MMHC, PC and FGES to the distributed data to learn the protein-DNA binding network. NOTEARS was excluded because its performance was not competitive. Local estimates of a competing method were combined to construct a global graph as we did in Section~\ref{sec:numerical}. To ease the comparison, we controlled the sparsity of estimated networks such that every method produced two graphs using the distributed data, with roughly $\widehat s_0= 17$ and 29 edges, except FGES which had difficulty generating output close to 17 edges. We also applied each method on the combined data (i.e, $K =1$) with the same parameters in Section~\ref{sec:numerical}. In this case, each estimated graph had around $\widehat s_0 = 30$ edges. The only exception is PC, whose estimate had 21 edges even when its significance level had been reduced to $10^{-10}$.

Since the true network structure is unknown, test data likelihood under multinomial DAG models in ten-fold cross validation is used to assess the accuracy of  estimated networks. Denote by $\gt$ and $g$ the likelihood values using training and test datasets, respectively, under multinomial DAG models (see supplemental material Section~\ref{sec:chip-likelihood} for calculation of $\gt$ and $g$). We also compute  $\text{BIC} = -2 \log  \gt + \log(\nt)  \calN(\widehat \calG)$ for model comparison, where $\nt$ is the training sample size and $\calN(\widehat \calG)$ is the number of multinomial parameters for estimated graph $\widehat\calG$. We choose some benchmarks to ease comparison. Denote by $g_{\text{B}}$ and $\text{BIC}_{\text{B}}$ the highest test data likelihood and the lowest BIC value, respectively. Define the log-likelihood difference $\Delta (\log g) = \log g - \log g_\text{B}$ and the BIC difference $\Delta \text{BIC} = \text{BIC} - \text{BIC}_\text{B}$. Note that since $\log g$ is test data log-likelihood while BIC is calculated with training data, the magnitude of $\Delta \text{BIC}$ is much larger than $\Delta (\log g)$. We also compute the value of $\exp{\{- \Delta \text{BIC} /(2\nt)\}}$ as an approximation to the normalized marginal likelihood ratio (NLR) $(P(\widetilde{X}\mid \widehat{\calG}) / P(\widetilde{X}\mid \widehat{\calG}_B))^{1/\nt}$, where $\widetilde{X}$ denotes training data, between estimated DAGs $\widehat{\calG}$ by a competing method and $\widehat{\calG}_B$ by the BIC benchmark. 

Table~\ref{tab:chip-delta-likelihood} summarizes $\widehat s_0$, $\Delta (\log g)$, $\Delta \text{BIC}$ and NLR of each method under three comparison settings, namely sparse, moderate and combined. The first two settings report results with different degree of sparsity in estimated graphs using distributed data over $K=20$ machines, and the last one shows results using combined data (i.e., $K=1$). In both sparse and moderate settings, DARLS achieves the highest test data likelihood and the smallest BIC, outperforming all the other methods by a substantial margin, which again demonstrates its effectiveness in DAG learning with distributed data. When applied to combined data, all methods, except PC, have higher test data likelihood than on distributed data, and the test data likelihood values are quite comparable. Comparing the likelihood between combined and distributed data for each method, we see that DARLS shows the smallest difference. In other words, among all the methods, DARLS has the smallest loss when applied to distributed data as compared to its result on the combined data.

It is worth mentioning that the likelihood for each method is calculated under the multinomial DAG model, instead of the GLDAG model. Thus, the superior performance of the distributed learning by DARLS on this real-world data suggests that our proposed GLDAG model is a good approximation to the underlying data generation mechanism. 

\begin{table}[t]
\centering
\caption{Comparison on the ChIP-Seq dataset. Best performance of each case is in bold.} 
\label{tab:chip-delta-likelihood}
\resizebox{\columnwidth}{!}{
\begin{tabular}{ccHcHccH
	@{\hskip 0.2in}HcHcHccH
	@{\hskip 0.2in}HcHcHccH}
\toprule
\multirow{2}{*}{Method} & \multicolumn{6}{c}{Sparse} & & & \multicolumn{6}{c}{Moderate} & & & \multicolumn{6}{c}{Combined}  \\
\cmidrule(r){2-24} 
&  $\widehat s_0$ &  par\_num & $\Delta (\log g)$ & $e^{\Delta g} - 1$ & $\Delta\text{BIC}$ & NLR & Complexity
& Method &  $\widehat s_0$ &  par\_num & $\Delta (\log g)$& $e^{\Delta g} - 1$& $\Delta\text{BIC}$ & NLR  & Complexity
& Method &  $\widehat s_0$ &  par\_num & $\Delta (\log g)$& $e^{\Delta g} - 1$& $\Delta\text{BIC}$ & NLR & Complexity \\ 
\cmidrule(r){2-7} 
\cmidrule(r){10-15} 
\cmidrule(r){18-23}
DARLS  &     16.5 &    35.0 &               $\mathbf{-136.9}$&                          $-$1.000 &   $\mathbf{1051.6}$ &                    0.501 &        0 
& DARLS  &     27.5 &    90.0 &                $\mathbf{-22.4}$ &                          $-$1.000 &    $\mathbf{297.7}$ &                    0.822 &        1 
& DARLS  &     31.0 &   138.0 &                 $-$3.7 &                          $-$0.976 &    540.2 &                    0.701 &        2 \\
MMHC   &     17.0 &    44.0 &               $-$149.6 &                          $-$1.000 &   2470.0 &                    0.198 &        0 
& MMHC   &     29.5 &   128.0 &                $-$44.4 &                          $-$1.000 &   1062.4 &                    0.498 &        1 
& MMHC   &     30.0 &    86.0 &                 $-$1.7 &                          $-$0.815 &      $\mathbf{0.0}$ &                    1.000 &        2 \\
FGES   &     11.5 &    24.0 &               $-$164.3 &                          $-$1.000 &   2478.5 &                    0.196 &        0 
& FGES   &     28.0 &   104.0 &                $-$39.6 &                          $-$1.000 &   1026.6 &                    0.510 &        1 
& FGES   &     34.0 &   122.0 &                 $-$2.7 &                          $-$0.935 &    250.9 &                    0.848 &        2 \\
HC     &     17.0 &    40.0 &               $-$151.9 &                          $-$1.000 &   2510.1 &                    0.192 &        0 
& HC     &     29.0 &   136.0 &                $-$36.8 &                          $-$1.000 &   1174.9 &                    0.462 &        1 
& HC     &     30.0 &    82.0 &                  $\mathbf{0.0}$ &                           0.000 &      7.9 &                    0.995 &        2 \\
PC     &     17.0 &    40.0 &               $-$156.3 &                          $-$1.000 &   2510.1 &                    0.192 &        0 
& PC     &     29.0 &   120.0 &                $-$36.9 &                          $-$1.000 &    942.6 &                    0.538 &        1 
& PC     &     21.0 &    52.0 &                $-$77.1 &                          $-$1.000 &   1047.8 &                    0.503 &        2 \\
\bottomrule
\end{tabular}
}
\end{table}

\begin{figure}
  \begin{center}
    \includegraphics[width=.8\textwidth]{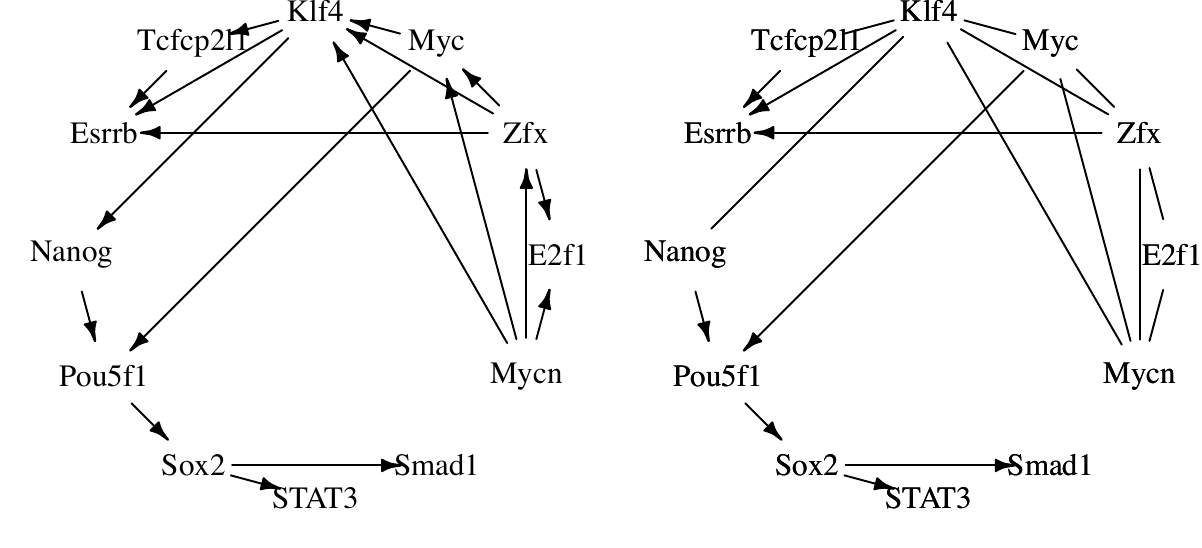}
  \end{center}
   \captionof{figure}{DAG (left) and the converted CPDAG (right) learned by DARLS.}
   \label{fig:DARLS-chip}
\end{figure}

To gain more scientific insights, we show in Figure~\ref{fig:DARLS-chip} the sparser DAG ($\widehat s_0=17$) and its converted CPDAG, learned by DARLS from the full dataset ($n = 8,462$) distributed over $K=20$ local machines, with $\lambda = 0.06$ and refinement parameter $\alpha = 0.3$. An interesting observation is the directed path {\em Nanog}$\to${\em Pou5f1}$\to${\em Sox2} in the estimated CPDAG, among the three core regulators in the gene regulatory network in mouse embryonic stem cells~\citep{Chen_etc_cell_08, Zetc07}. It is well-known that many genes are co-regulated by {\em Pou5f1}, {\em Sox2} and {\em Nanog}. The estimated path suggests that {\em Nanog} binding would cause the binding of {\em Pou5f1}, which then may cause {\em Sox2} binding. This provides new clue for how the three TFs work together to co-regulate downstream genes. Data analysis in \cite{Chen_etc_cell_08}, the original work that generated the ChIP-Seq data, suggests that there are two clusters of TFs that tend to co-bind: The first group consists of {\em Nanog}, {\em Sox2}, {\em Oct4}, {\em Smad1} and {\em STAT3}, while the second group includes {\em Mycn}, {\em Myc}, {\em Zfx} and {\em E2f1}. These two groups are clearly recovered in the estimated CPDAG, which contains a dense undirected subgraph on the second group of TFs and a fully directed subgraph on the first group. Moreover, the directed edge {\em Myc}$\to${\em Pou5f1} indicates that the second group might be in the causal upstream of the first group, a novel hypothesis for potential experimental investigation.


\section{Discussion}
\label{sec:conclude}

In this paper, we develop the DARLS algorithm that incorporates a distributed optimization method in simulated annealing to learn causal graphs from distributed data. Based on simulation studies and a real data application, we have shown that DARLS is highly competitive even when its model assumptions are violated. In its distributed optimization given an ordering, DARLS learns a causal graph by optimizing a convex penalized likelihood. In practice, one may consider concave penalties~\citep{AZ15, YAZ19} to improve accuracy when learning DAGs, although there may be lack of theoretical guarantees for convergence of distributed learning with a concave penalty. This is certainly a promising further development for our method.

Our proposed GLDAG model includes a family of flexible distributions besides linear Gaussian models (with equal variance), and thus can be applied to different types of data. It is also possible to add interaction terms to GLDAGs~\eqref{eq:expo-fam-def} to model nonlinear effects between variables. Such generalization is expected to approximate real causal relations with higher accuracy. We have established that continuous GLDAGs are identifiable, justifying their use in causal discovery, and it is left as future work to study the identifiability of general GLDAGs.

The primary focus of this paper is on big distributed data, with large $n$ but moderate $p$. In this setting, we established the convergence of the solution obtained by distributed optimization to a global minimizer of the loss using the combined data, and the consistency of the global minimizer as an estimate of the true DAG parameter. However, generalizing the convergence and consistency results to allow diverging $p$ is theoretically interesting and left as future work.


\bibliographystyle{asa}
\bibliography{cite2}

\begin{thebibliography}{53}
\newcommand{\enquote}[1]{``#1''}
\expandafter\ifx\csname natexlab\endcsname\relax\def\natexlab#1{#1}\fi

\bibitem[{Apple and Google(2021)}]{ENPA}
Apple and Google (2021), \enquote{Explosure {N}otification {P}rivacy-preserving
  {A}nalytics (ENPA) White Paper,} Tech. rep., Apple and Google.

\bibitem[{Aragam et~al.(2019)Aragam, Gu, and Zhou}]{AGZ19}
Aragam, B., Gu, J., and Zhou, Q. (2019), \enquote{Learning {L}arge-{S}cale
  {B}ayesian {N}etworks with the sparsebn {P}ackage,} \textit{Journal of
  Statistical Software}, 91, 1--38.

\bibitem[{Aragam and Zhou(2015)}]{AZ15}
Aragam, B. and Zhou, Q. (2015), \enquote{Concave {P}enalized {E}stimation of
  {S}parse {G}aussian {B}ayesian {N}etworks,} \textit{Journal of Machine
  Learning Research}, 16, 2273--2328.

\bibitem[{Cai et~al.(2020)Cai, Kang, and Yu}]{C2020}
Cai, Q., Kang, J., and Yu, T. (2020), \enquote{Bayesian {N}etwork {M}arker
  {S}election via the {T}hresholded {G}raph {L}aplacian {G}aussian {P}rior,}
  \textit{Bayesian Analysis}, 15, 79--102.

\bibitem[{Chen et~al.(2008)Chen, Xu, Yuan, Fang, Huss, Vega, Wong, Orlov,
  Zhang, Jiang, Loh, Yeo, Yeo, Narang, Govindarajan, Leong, Shahab, Ruan,
  Bourque, Sung, Clarke, Wei, and Ng}]{Chen_etc_cell_08}
Chen, X., Xu, H., Yuan, P., Fang, F., Huss, M., Vega, V.~B., Wong, E., Orlov,
  Y.~L., Zhang, W., Jiang, J., Loh, Y.-H., Yeo, H.~C., Yeo, Z.~X., Narang, V.,
  Govindarajan, K.~R., Leong, B., Shahab, A., Ruan, Y., Bourque, G., Sung,
  W.-K., Clarke, N.~D., Wei, C.-L., and Ng, H.-H. (2008), \enquote{Integration
  of {E}xternal {S}ignaling {P}athways with the {C}ore {T}ranscriptional
  {N}etwork in {E}mbryonic {S}tem {C}ells,} \textit{Cell}, 133, 1106--17.

\bibitem[{Chickering(2002)}]{C02}
Chickering, D.~M. (2002), \enquote{Optimal {S}tructure {I}dentification with
  {G}reedy {S}earch,} \textit{Journal of Machine Learning Research}, 3,
  507--554.

\bibitem[{Dehejia and Wahba(1999)}]{DW99}
Dehejia, R.~H. and Wahba, S. (1999), \enquote{Causal {E}ffects in
  {N}onexperimental {S}tudies: {R}eevaluating the {E}valuation of {T}raining
  {P}rograms,} \textit{Journal of the American Statistical Association}, 94,
  1053--1062.

\bibitem[{Fan et~al.(2019)Fan, Guo, and Wang}]{FGW19}
Fan, J., Guo, Y., and Wang, K. (2019), \enquote{Communication-Efficient
  Accurate Statistical Estimation,} \textit{arXiv:1906.04870}.

\bibitem[{Friedman and Koller(2003)}]{FK03}
Friedman, N. and Koller, D. (2003), \enquote{{B}eing {B}ayesian about {N}etwork
  {S}tructure. {A} {B}ayesian {A}pproach to {S}tructure {D}iscovery in
  {B}ayesian {N}etworks,} \textit{Machine Learning}, 50, 95--125.

\bibitem[{Fu and Zhou(2013)}]{FZ13}
Fu, F. and Zhou, Q. (2013), \enquote{Learning {S}parse {C}ausal {G}aussian
  {N}etworks with {E}xperimental {I}ntervention: {R}egularization and
  {C}oordinate {D}escent,} \textit{Journal of the American Statistical
  Association}, 108, 288--300.

\bibitem[{G\'{a}mez et~al.(2011)G\'{a}mez, Mateo, and Puerta}]{GMP11}
G\'{a}mez, J.~A., Mateo, J.~L., and Puerta, J.~M. (2011), \enquote{Learning
  {B}ayesian {N}etworks by {H}ill {C}limbing: {E}fficient {M}ethods {B}ased on
  {P}rogressive {R}estriction of the {N}eighborhood,} \textit{Data Mining and
  Knowledge Discovery}, 22, 106--148.

\bibitem[{Glymour(2006)}]{M06}
Glymour, M.~M. (2006), \enquote{Using {C}ausal {D}iagrams to {U}nderstand
  {C}ommon {P}roblems in {S}ocial {E}pidemiology,} in \textit{Methods in Social
  Epidemiology}, John Wiley and Sons, pp. 393--428.

\bibitem[{Gou et~al.(2007)Gou, Jun, and Zhao}]{GJZ10}
Gou, K.~X., Jun, G.~X., and Zhao, Z. (2007), \enquote{Learning {B}ayesian
  {N}etwork {S}tructure from {D}istributed {H}omogeneous {D}ata,}
  \textit{Eighth ACIS International Conference on Software Engineering,
  Artificial Intelligence, Networking, and Parallel/Distributed Computing}, 3,
  250--254.

\bibitem[{Gu et~al.(2019)Gu, Fu, and Zhou}]{GFZ18}
Gu, J., Fu, F., and Zhou, Q. (2019), \enquote{Penalized {E}stimation of
  {D}irected {A}cyclic {G}raphs from {D}iscrete {D}ata,} \textit{Statistics and
  Computing}, 29, 161--176.

\bibitem[{Guo et~al.(2020)Guo, Cheng, Li, Hahn, and Liu}]{Guo_etc_20}
Guo, R., Cheng, L., Li, J., Hahn, P.~R., and Liu, H. (2020), \enquote{A
  {S}urvey of {L}earning {C}ausality with {D}ata: {P}roblems and {M}ethods,}
  \textit{ACM Computing Surveys}, 53, 1--37.

\bibitem[{Heckerman et~al.(1995)Heckerman, Geiger, and Chickering}]{HGC95}
Heckerman, D., Geiger, D., and Chickering, D.~M. (1995), \enquote{Learning
  {B}ayesian {N}etworks: {T}he {C}ombination of {K}nowledge and {S}tatistical
  {D}ata,} \textit{Machine Learning}, 20, 197--243.

\bibitem[{Hill(2012)}]{Hill12}
Hill, J.~L. (2012), \enquote{Bayesian {N}onparametric {M}odeling for {C}ausal
  {I}nference,} \textit{Journal of Computational and Graphical Statistics}, 20,
  217--240.

\bibitem[{Hoyer et~al.(2008)Hoyer, Janzing, Mooij, Peters, and
  Sch\"olkopf}]{HJMPS08}
Hoyer, P.~O., Janzing, D., Mooij, J., Peters, J., and Sch\"olkopf, B. (2008),
  \enquote{Nonlinear {C}ausal {D}iscovery with {A}dditive {N}oise {M}odels,}
  \textit{Preceedings of 21st International Conference on Neural Information
  Processing Systems}, 689--696.

\bibitem[{Imbens(2004)}]{Imbens04}
Imbens, G.~W. (2004), \enquote{Nonparametric {E}stimation of {A}verage
  {T}reatment {E}ffects under {E}xogeneity: {A} {R}eview,} \textit{The Review
  of Economics and Statistics}, 86, 4--29.

\bibitem[{Jordan et~al.(2018)Jordan, Lee, and Yang}]{JLY18}
Jordan, M.~I., Lee, J.~D., and Yang, Y. (2018),
  \enquote{Communication-{E}fficient {D}istributed {S}tatistical {I}nference,}
  \textit{Journal of the American Statistical Association}, 114, 668--681.

\bibitem[{Kalisch et~al.(2012)Kalisch, M\''{a}chler, Colombo, Maathuis, and
  B\''{u}hlmann}]{Metc12}
Kalisch, M., M\''{a}chler, M., Colombo, D., Maathuis, M.~H., and B\''{u}hlmann,
  P. (2012), \enquote{Causal {I}nference {U}sing {G}raphical {M}odels with the
  {R} {P}ackage pcalg,} \textit{Journal of Statistical Software}, 47, 1--26.

\bibitem[{Larra\~{n}aga et~al.(1996)Larra\~{n}aga, Poza, Yurramendi, Murga, and
  Kuijpers}]{Larranaga1996}
Larra\~{n}aga, P., Poza, M., Yurramendi, Y., Murga, R.~H., and Kuijpers, C.
  M.~H. (1996), \enquote{Structure {L}earning of {B}ayesian {N}etworks by
  {G}enetic {A}lgorithms: {A} {P}erformance {A}nalysis of {C}ontrol
  {P}arameters,} \textit{IEEE Transactions on Pattern Analysis and Machine
  Intelligence}, 18, 912--926.

\bibitem[{Lee and Kim(2010)}]{L2010}
Lee, N. and Kim, J.-M. (2010), \enquote{Conversion of {C}ategorical {V}ariables
  into {N}umerical {V}ariables via {B}ayesian {N}etwork {C}lassifiers for
  {B}inary {C}lassifications,} \textit{Computational Statistics {Harold \&
  Maude} Data Analysis}, 54, 1247--1265.

\bibitem[{Meek(1995)}]{M1995}
Meek, C. (1995), \enquote{Strong {C}ompleteness and {F}aithfulness in
  {B}ayesian {N}etworks,} \textit{Proceedings of the Eleventh conference on
  Uncertainty in artificial intelligence}, 411--418.

\bibitem[{Mehmood et~al.(2016)Mehmood, Natgunanathan, Xiong, Hua, and
  Guo}]{Mehmood_etc_16}
Mehmood, A., Natgunanathan, I., Xiong, Y., Hua, G., and Guo, S. (2016),
  \enquote{Protection of {B}ig {D}ata {P}rivacy,} \textit{IEEE Access}, 4,
  1821--1834.

\bibitem[{Molzahn et~al.(2017)Molzahn, D{\"o}rfler, Sandberg, Low, Chakrabarti,
  Baldick, and Lavaei}]{Metc17}
Molzahn, D.~K., D{\"o}rfler, F., Sandberg, H., Low, S.~H., Chakrabarti, S.,
  Baldick, R., and Lavaei, J. (2017), \enquote{A {S}urvey of {D}istributed
  {O}ptimization and {C}ontrol {A}lgorithms for {E}lectric {P}ower {S}ystems,}
  \textit{IEEE Transactions on Smart Grid}, 8, 2941--2962.

\bibitem[{Na and Yang(2010)}]{NY10}
Na, Y. and Yang, J. (2010), \enquote{{D}istributed {B}ayesian {N}etwork
  {S}tructure {L}earning,} \textit{IEEE International Symposium on Industrial
  Electronics}, 1607--1611.

\bibitem[{Ouyang et~al.(2009)Ouyang, Zhou, and Wong}]{OZW09}
Ouyang, Z., Zhou, Q., and Wong, W.~H. (2009), \enquote{ChIP-{S}eq of
  {T}ranscription {F}actors {P}redicts {A}bsolute and {D}ifferential {G}ene
  {E}xpression in {E}mbryonic {S}tem {C}ells,} \textit{Preceedings of the
  National Academic of Science of the United State of America}, 106, 21521--6.

\bibitem[{Parikh and Boyd(2013)}]{NP13}
Parikh, N. and Boyd, S. (2013), \enquote{Proximal Algorithms,}
  \textit{Foundations and Trends in Optimization}, 1, 123--231.

\bibitem[{Pearl(1995)}]{Pearl95}
Pearl, J. (1995), \enquote{Causal {D}iagrams for {E}mpirical {R}esearch,}
  \textit{Biometrika}, 82, 669--710.

\bibitem[{Pearl(2009)}]{Pearl09}
--- (2009), \enquote{Causal {I}nference in {S}tatistics: {A}n {O}verview,}
  \textit{Statistical Surveys}, 3, 96--146.

\bibitem[{Peters and B\"{u}hlmann(2014)}]{PB14}
Peters, J. and B\"{u}hlmann, P. (2014), \enquote{Identifiability of {G}aussian
  {S}tructural {M}odels with {E}qual {E}rror {V}ariances,} \textit{Biometrika},
  101, 219--228.

\bibitem[{Peters et~al.(2014)Peters, Mooij, Janzing, and
  Sch\"olkopf}]{PMJS14-JMLR}
Peters, J., Mooij, J.~M., Janzing, D., and Sch\"olkopf, B. (2014),
  \enquote{Causal {D}iscovery with {C}ontinuous {A}dditive {N}oise {M}odels,}
  \textit{Journal of Machine Learning Research}, 15, 2009--2053.

\bibitem[{Ramanan and Natarajan(2020)}]{QCK20}
Ramanan, N. and Natarajan, S. (2020), \enquote{Causal {L}earning from
  {P}redictive {M}odeling for {O}bservational {D}ata,} \textit{Frontiers in Big
  Data}, 3, 535976.

\bibitem[{Ramsey et~al.(2017)Ramsey, Glymour, Sanchez-Romero, and
  Glymour}]{R17}
Ramsey, J., Glymour, M., Sanchez-Romero, R., and Glymour, C. (2017), \enquote{A
  {M}illion {V}ariables and {M}ore: the {F}ast {G}reedy {E}quivalence {S}earch
  {A}lgorithm for {L}earning {H}igh-dimensional {G}raphical {C}ausal {M}odels,
  with an {A}pplication to {F}unctional {M}agnetic {R}esonance {I}mages,}
  \textit{International Journal of Data Science and Analytics}, 3, 121--129.

\bibitem[{Scanagatta et~al.(2015)Scanagatta, de~Campos, Corani, and
  Zaffalon}]{SCC15}
Scanagatta, M., de~Campos, C.~P., Corani, G., and Zaffalon, M. (2015),
  \enquote{Learning {B}ayesian {N}etworks with {T}housands of {V}ariables,}
  \textit{{\normalfont in} Advances in Neural Information Processing Systems},
  1864--1872.

\bibitem[{Scutari(2007)}]{bnrep}
Scutari, M. (2007), \enquote{Bayesian Network Repository,}
  \url{http://www.bnlearn.com/bnrepository/}, accessed: 2020-05-01.

\bibitem[{Scutari(2010)}]{S10}
--- (2010), \enquote{Learning {B}ayesian {N}etworks with the bnlearn {R}
  {P}ackage,} \textit{Journal of Statistical Software}, 35, 1--22.

\bibitem[{Shamir et~al.(2014)Shamir, Srebro, and Zhang}]{SSZ14}
Shamir, O., Srebro, N., and Zhang, T. (2014),
  \enquote{Communication-{E}fficient {D}istributed {O}ptimization using an
  {A}pproximate {N}ewton-type {M}ethod,} \textit{Proceedings of the 31st
  International Conference on International Conference on Machine Learnings},
  32, 1000--1008.

\bibitem[{Shimizu et~al.(2006)Shimizu, Hoyer, Hyv\"arinen, and
  Kerminen}]{lingam06}
Shimizu, S., Hoyer, P.~O., Hyv\"arinen, A., and Kerminen, A. (2006), \enquote{A
  {L}inear {N}on-{G}aussian {A}cyclic {M}odel for {C}ausal {D}iscovery,}
  \textit{Journal of Machine Learning Research}, 7, 2003--2030.

\bibitem[{Song et~al.(2021)Song, Zhou, Kang, Aung, Zhang, Zhao, Needham,
  Kardia, Liu, Meeker, Smith, and Mukherjee}]{Setc2021}
Song, Y., Zhou, X., Kang, J., Aung, M.~T., Zhang, M., Zhao, W., Needham, B.~L.,
  Kardia, S. L.~R., Liu, Y., Meeker, J.~D., Smith, J.~A., and Mukherjee, B.
  (2021), \enquote{Bayesian {S}parse {M}ediation {A}nalysis with {T}argeted
  {P}enalization of {N}atural {I}ndirect {E}ffects,} \textit{Journal of the
  Royal Statistical Society. Series C, Applied statistics}, 70, 1391--1412.

\bibitem[{Spirtes and Glymour(1991)}]{pc_91}
Spirtes, P. and Glymour, C. (1991), \enquote{An {A}lgorithm for {F}ast
  {R}ecovery of {S}parse {C}ausal {G}raphs,} \textit{Social Science Computer
  Review}, 9, 62--72.

\bibitem[{Tang et~al.(2019)Tang, Wang, Nguyen, and Altintas}]{TWNA19}
Tang, Y., Wang, J., Nguyen, M., and Altintas, I. (2019), \enquote{{PE}n{B}ayes:
  {A} {M}ulti-Layered {E}nsemble {A}pproach for {L}earning {B}ayesian {N}etwork
  {S}tructure from {B}ig {D}ata,} \textit{Sensors}, 19, 4400.

\bibitem[{Tsamardinos et~al.(2006)Tsamardinos, Brown, and Aliferis}]{TEA06}
Tsamardinos, I., Brown, L.~E., and Aliferis, C.~F. (2006), \enquote{The
  {M}ax-{M}in {H}ill-{C}limbing {B}ayesian {N}etwork {S}tructure {L}earning
  {A}lgorithm,} \textit{Machine Learning}, 65, 31--78.

\bibitem[{Wang and Zhou(2021)}]{WZ21}
Wang, B. and Zhou, Q. (2021), \enquote{Causal {N}etwork {L}earning with
  {N}0n-intertible {F}unctional {R}elationships,} \textit{Computational
  Statistics and Data Analysis}, 156, 107141.

\bibitem[{Yang et~al.(2019)Yang, Yi, Wu, Yuan, Wu, Meng, Hong, Wang, Lin, and
  Johansson}]{YANG19}
Yang, T., Yi, X., Wu, J., Yuan, Y., Wu, D., Meng, Z., Hong, Y., Wang, H., Lin,
  Z., and Johansson, K.~H. (2019), \enquote{A {S}urvey of {D}istributed
  {O}ptimization,} \textit{Annual Reviews in Control}, 47, 278--305.

\bibitem[{Ye et~al.(2021)Ye, Amini, and Zhou}]{YAZ19}
Ye, Q., Amini, A.~A., and Zhou, Q. (2021), \enquote{Optimizing {R}egularized
  {C}holesky {S}core for {O}rder-{B}ased {L}earning of {B}ayesian {N}etworks,}
  \textit{IEEE Transactions on Pattern Analysis \& Machine Intelligence}, 43,
  3555--3572.

\bibitem[{Yuan and Lin(2007)}]{YL07}
Yuan, M. and Lin, Y. (2007), \enquote{Model {S}election and {E}stimation in
  {R}egression with {G}rouped {V}ariables,} \textit{Journal of Royal
  Statistical Society, Series B}, 68, 49--67.

\bibitem[{Zhang et~al.(2013)Zhang, Duchi, and Wainwright}]{ZDW13}
Zhang, Y., Duchi, J.~C., and Wainwright, M.~J. (2013),
  \enquote{{C}ommunication-{E}fficient {A}lgorithms for {S}tatistical
  {O}ptimization,} \textit{Journal of Machine Learning Research}, 14,
  3321--3363.

\bibitem[{Zheng(2018)}]{notears_codes}
Zheng, X. (2018), \enquote{DAGs with NO TEARS,}
  \url{https://github.com/xunzheng/notearss}, accessed: 2021-09-01.

\bibitem[{Zheng et~al.(2018)Zheng, Aragam, Pavikumar, and Xing}]{ZA18-NOTEARS}
Zheng, X., Aragam, B., Pavikumar, P., and Xing, E.~P. (2018), \enquote{DAGs
  with {NO} {TEARS}: {C}ontinuous {O}ptimization for {S}tructure {L}earning,}
  \textit{{\normalfont in} Advances in Neural Information Processing Systems}.

\bibitem[{Zhou et~al.(2007)Zhou, Chipperfield, Melton, and Wong}]{Zetc07}
Zhou, Q., Chipperfield, H., Melton, D.~A., and Wong, W.~H. (2007), \enquote{A
  {G}ene {R}egulatory {N}etwork in {M}ouse {E}mbryonic {S}tem {C}ells,}
  \textit{Preceedings of the National Academy of Sciences of the United States
  of America}, 104, 16438--16443.

\bibitem[{Zinkevich et~al.(2010)Zinkevich, Weimer, Smola, and Li}]{ZWL10}
Zinkevich, M.~A., Weimer, M., Smola, A., and Li, L. (2010),
  \enquote{Parallelized {S}tochastic {G}radient {D}escent,} in \textit{Advances
  in Neural Information Processing Systems}, vol.~23, pp. 2595--2603.

\end{thebibliography}

\end{document}